\documentclass[useAMS,usenatbib]{mn2e}

\usepackage{graphicx}
\usepackage{amsmath}

\def\ba{\begin{eqnarray}}
\def\ea{\end{eqnarray}}
\def\be{\begin{equation}}
\def\ee{\end{equation}}
\def\K{\text{\textbf{\textit{K}}}}

\usepackage{macros}
\usepackage[usenames,dvipsnames,svgnames,table]{xcolor}
\usepackage{hyperref}
\definecolor{darkblue}{rgb}{0.0,0.0,0.3}
\hypersetup{colorlinks,
            linkcolor=darkblue,urlcolor=darkblue,
            anchorcolor=darkblue,citecolor=darkblue}

\usepackage{txfonts}

\DeclareSymbolFont{cmletters}{OML}{cmm}{m}{it}
\DeclareMathSymbol{v}{\mathalpha}{cmletters}{"76}
\title[Evolution of non-spherical pulsars]{Evolution of
  non-spherical pulsars with plasma-filled magnetospheres}
\author[Lev Arzamasskiy, Alexander Philippov, Alexander Tchekhovskoy]{Lev Arzamasskiy$^{1}$\thanks{E-mail:
lev.arzamasskiy@phystech.edu}, Alexander Philippov$^{2}$, Alexander Tchekhovskoy$^{3, 4}$\\
$^{1}$Moscow Institute of Physics and Technology, Dolgoprudny, Institutsky per., 9, Moscow region, 141700, Russia\\
$^{2}$Department of Astrophysical Sciences, Peyton Hall, Princeton University, Princeton, NJ 08544, USA\\
$^{3}$Departments of Astronomy and Physics, Theoretical Astrophysics
Center, University of California Berkeley, Berkeley, CA 94720-3411\\
$^4$Lawrence Berkeley National Laboratory, 1 Cyclotron Rd, Berkeley, CA 94720, USA}
\begin{document}

\date{Accepted. Received; in original form}
\pagerange{\pageref{firstpage}--\pageref{lastpage}} \pubyear{2015}

\maketitle

\label{firstpage}

\begin{abstract}
  Pulsars are famous for their rotational stability. Most of them
  steadily spin down and display a highly repetitive pulse shape. But
  some pulsars experience timing irregularities such as nulling,
  intermittency, mode changing and timing noise. As changes in the
  pulse shape are often correlated with timing irregularities,
  precession is a possible cause of these phenomena. Whereas pulsar
  magnetospheres are filled with plasma, most pulsar precession
  studies were carried out within the vacuum approximation and
  neglected the effects of magnetospheric currents and charges. Recent
  numerical simulations of plasma-filled pulsar magnetospheres provide
  us with a detailed quantitative description of magnetospheric
  torques exerted on the pulsar surface.  In this paper, we present
  the study of neutron star evolution using these new torque
  expressions. We show that they lead to (1) much slower long-term
  evolution of pulsar parameters and (2) much less extreme solutions
  for these parameters than the vacuum magnetosphere models.  To
  facilitate the interpretation of observed pulsar timing residuals,
  we derive an analytic model that (1) describes the time evolution of
  non-spherical pulsars and (2) translates the observed pulsar timing
  residuals into the geometrical parameters of the pulsar. We apply
  this model to two pulsars with very different temporal behaviours.
  For the pulsar B1828-11, we demonstrate that the timing residual
  curves allow two pulsar geometries: one with stellar deformation
  pointing along the magnetic axis and one along the rotational
  axis. For the Crab pulsar, we use the model show that the recent
  observation of its magnetic and rotational axes moving away from
  each other can be explained by
  precession. 
\end{abstract}

\begin{keywords}
stars: magnetic field-- stars: neutron -- pulsars: general -- stars: rotation.
\end{keywords}
\section{Introduction}
\label{sect:intro}

Studies of pulsar evolution are often performed under the assumption
of a spherical neutron star in vacuum. In addition to such idealized
studies \citep{deutsch55}, it is now possible to perform more
realistic numerical simulations in the limit of a plasma-filled
magnetosphere \citep{ckf99, spit06,kc09, SashaMHD}, and more recently
in the full kinetic limit \citep{2014ApJ...785L..33P, ChenPIC,
  2015MNRAS.448..606C, 2015ApJ...801L..19P,
  2015NewA...36...37B}. These models predict the evolution of the
pulsar rotation period $P$ and the inclination angle $\alpha$ that the
magnetic axis makes with the rotational axes. For a
spherically-symmetric star, the pulsar period slowly increases and the
inclination angle gradually decreases towards $0^\circ$
(\citealt{ptl14}; but see \citealt{BGI}). However, deviations in the
pulsar pulse arrival times from the expectations of the above smooth
time-evolution, e.g., pulsar timing noise and
intermittency, indicate that pulsar spindown may experience quite
different, often periodical changes.

One way to introduce an oscillatory behaviour into the spindown is to
consider a non-spherical star. In the absence of a physically motivated
model of a plasma-filled magnetosphere, previous studies \citep{G70, Melatos} considered pulsar evolution in the vacuum
approximation. In this paper, we investigate the evolution of non-spherical pulsars whose
magnetospheres are filled with plasma. For this, we use recent results of
magnetohydrodynamic (MHD) simulations of plasma-filled magnetospheres \citep{ptl14}.

The paper is organized as follows. 
We start with the discussion of the vacuum approximation for pulsar spindown in Section~\ref{sect:vacuum}. 
We discuss the parametrisation of magnetospheric torques in
Section~\ref{sect:plasma}. The solutions for spherical pulsars are
presented in Section~\ref{sect:sphere}. The effects of stellar non-sphericity
are discussed in Section~\ref{sect:precession}. 
We discuss the implications of our results in
Section~\ref{sect:conclusions} and conclude in Section~\ref{sec:conclusions}.

\section{Vacuum solution}
\label{sect:vacuum}

A pulsar rotating in vacuum emits magnetodipole
radiation that carries away the angular momentum of the neutron star.
The resulting torque acting on the star could be found by
integrating the stresses exerted at its surface,
\begin{equation}
K_i = \int \epsilon_{ijk} r_j n_l T_{kl}{\rm d}S,
\label{eq:torque}
\end{equation}
where the summation over repeated indices is implied, $T_{kl}$ is the
stress tensor (in our case, we have only electromagnetic forces acting
on the star, so it is Maxwell's stress tensor), $\epsilon_{ijk}$ is the
Levi-Civita symbol, ${\boldsymbol r}$ is the position of surface
element with an area element ${\rm d} S$, and $n_l$ is the normal vector.

If we know electric and magnetic field components near the stellar surface, it is
straightforward to compute the torque $\boldsymbol K$ from
eq.~(\ref{eq:torque}). \citet{deutsch55} computed vacuum electromagnetic
fields near a rotating spherical star endowed with a dipolar
magnetic field. Using these expressions, \citet{Michel70}
derived the torques acting on the pulsar in vacuum. If pulsar's angular frequency
vector ${\boldsymbol \Omega}$
is oriented along the $z$-axis and the pulsar magnetic moment vector 
$\boldsymbol\mu$ lies in the $x$ -- $z$ plane, the torque components
in the vacuum limit  are:
\begin{align}
K_z &= -\frac{2}{3} K_{\rm aligned} \sin^2 \alpha, \label{eq:Kzvac}\\
K_x &= \phantom{-}\frac{2}{3} K_{\rm aligned} \sin \alpha \cos \alpha, \label{eq:Kxvac}
\end{align}
where 
\begin{equation}
K_{\rm aligned} = \frac{\mu^2 \Omega^3}{c^3}
\end{equation}
and $\alpha$ is the
inclination angle, i.e., the angle between $\boldsymbol \Omega$ and~$\boldsymbol \mu$.

The torque component due to eq.~\eqref{eq:Kzvac} slows down stellar
rotation. The torque component given by eq.~\eqref{eq:Kxvac} points
towards $\boldsymbol \mu$.  Under the action of this torque component,
the inclination angle $\alpha$ evolves toward the alignment, and the
star approaches
the state of minimum angular momentum and energy
loss \citep{ptl14}. 

Though the vacuum magnetosphere model predicts the spindown rate close to
the observed value, it has several important shortcomings. The
magnetospheres of real pulsars are filled with plasma that
inevitably modifies the structure of the magnetic fields via the
charges and currents that it carries \citep{GJ}. We discuss these plasma effects below.

\section{Magnetospheric torques}
\label{sect:plasma}

Unfortunately, there is no simple analytical solution for
magnetospheric torques in the plasma-filled magnetosphere case. Thus,
we will rely on the recent advances in numerical
simulations of oblique pulsar magnetospheres \citep{spit06,kc09,
  SashaMHD,2015ApJ...801L..19P,2015arXiv150301467T} that allow us to quantitatively study the influence
of magnetospheric charges and currents on the structure of the
magnetosphere as well as on the pulsar spindown and alignment.

\citet{ptl14} analyzed the results of force-free and MHD simulations and came up with
the following parametrization of magnetospheric torques for
\emph{plasma-filled magnetospheres}, to which for simplicity we will
refer to below as simply \emph{MHD magnetospheres}:
\begin{align}
\label{eqn:Kz}
K_z &= - K_{\rm aligned} (k_0 + k_1 \sin^2 \alpha),\\
\label{eqn:Kx}
K_x &= k_2 K_{\rm aligned} \sin \alpha \cos \alpha,
\end{align}
where $k_0$, $k_1$ and $k_2$ are factors of order unity. They
weakly depend on $R_*/R_{LC}$, the ratio of the stellar radius $R_*$ to the light cylinder radius $R_{LC} = c/\Omega$. Typical parameters for vacuum and
MHD models are listed in Table~\ref{table:models} \citep{ptl14}.

In addition to torque components (\ref{eqn:Kz}) and (\ref{eqn:Kx}), we
need to know the anomalous torque, or the torque component directed along ${\boldsymbol \Omega} \times
{\boldsymbol \mu}$, which could be written as
\begin{equation}
\label{eqn:Ky}
K_y = k_3 K_{\rm aligned} \frac{c}{\Omega R}\sin\alpha\cos\alpha,
\end{equation}
where $k_3 = {\rm const}$. \citet{Melatos} estimated $k_3=0.3$. In numerical simulations, anomalous torque is hard to
measure as it diverges with $R \rightarrow 0$. We estimate it in our MHD model to be
$k_3 \approx 0.1$ (please see Table \ref{table:models}). As we will show below, the
  timing residuals \emph{do not depend} on the strength of the anomalous
  torque and the value of $k_3$ (see Section \ref{sect:analytics}).

\begin{table}
\centering
\caption{Spindown parameters of different magnetospheric models. $k_0$, $k_1$ and $k_2$ are defined in equations (\ref{eqn:Kz}) - (\ref{eqn:Kx}), and $k_3$ is referred to the anomalous torque and is defined in equation (\ref{eqn:Ky}).}
\begin{tabular}{l|c|c|c|c}
\hline
Model & $k_0$ & $k_1$ & $k_2$ & $k_3$\\
\hline
\hline
Vacuum & 0 & 2/3 & 2/3 & $0.3$\\
MHD/force-free & 1 & 1 & 1 & $\sim0.1$\\
\hline
\end{tabular}
\label{table:models}
\end{table}

The most important difference between vacuum and
MHD models is in the value of parameter $k_0$, which describes the energy losses of an
aligned rotator.
In vacuum, an aligned rotator does not spin down, i.e. $\alpha = 0$, $\Omega
\ne 0$ is a solution to evolution equations. This is impossible in the
MHD case, because the spindown luminosity of an  aligned
rotator is non-zero.

\section{Evolution of spherical stars}
\label{sect:sphere}

\begin{figure*}
\centering
\includegraphics[scale=0.6]{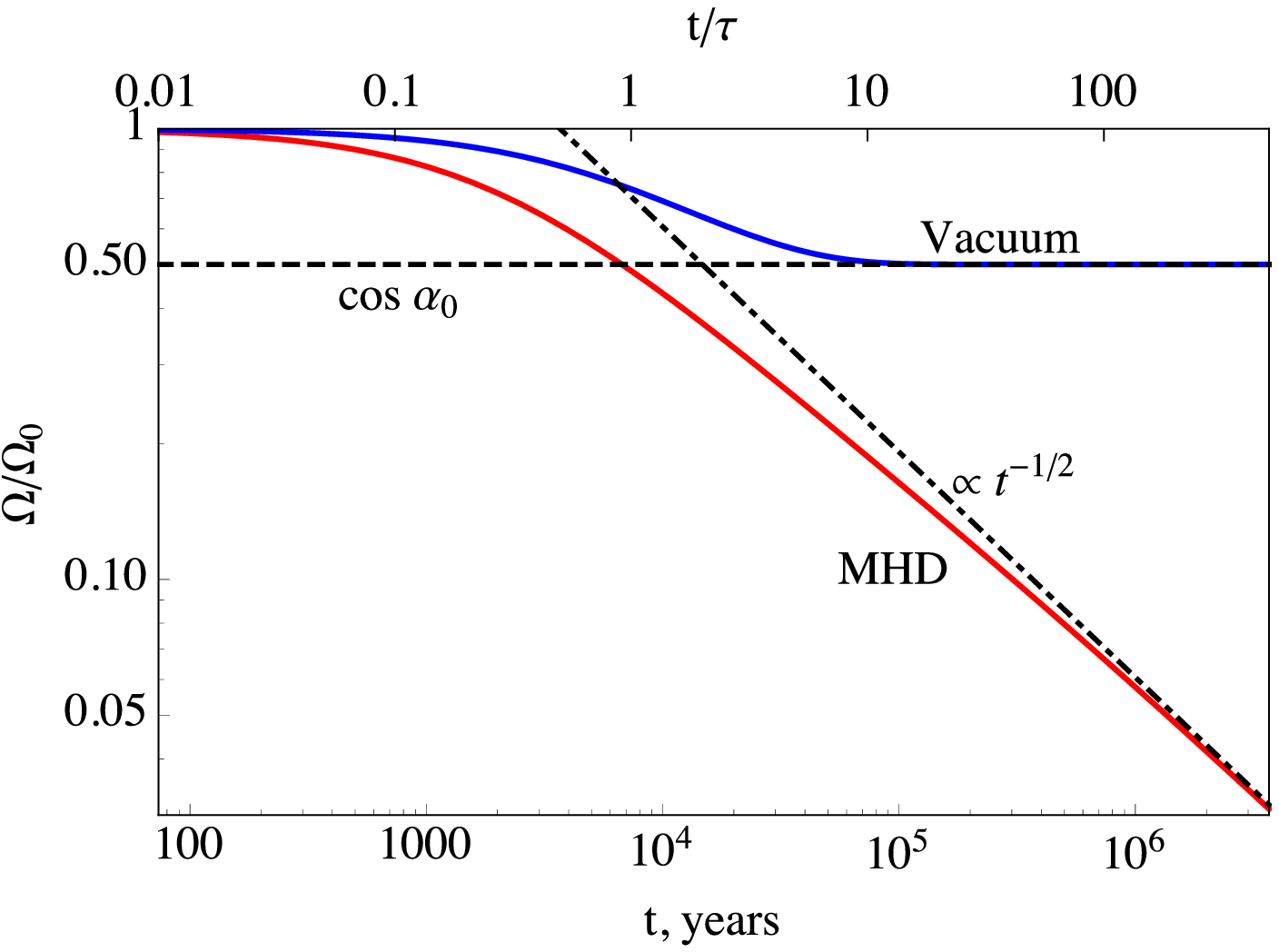}~~~\includegraphics[scale=0.6]{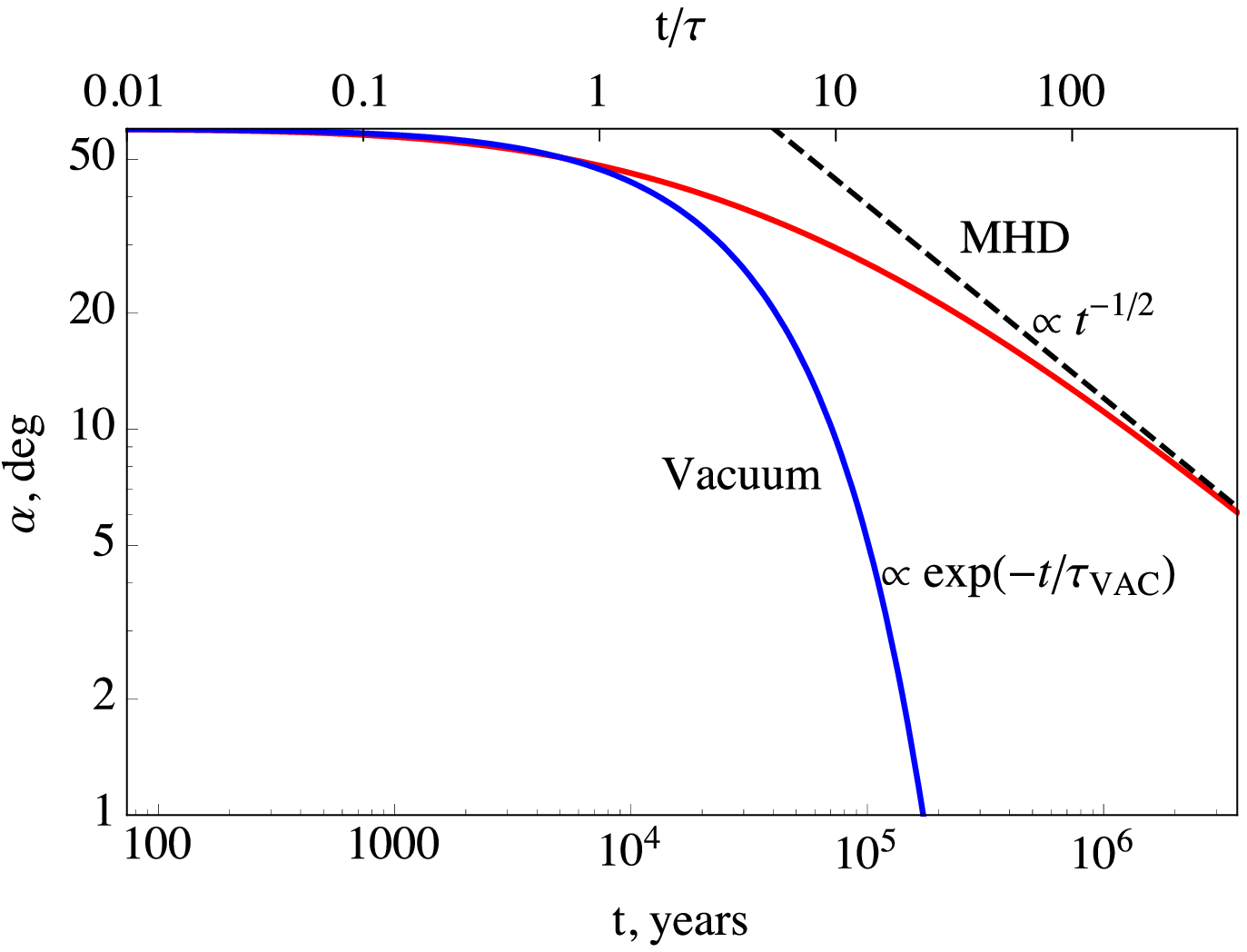}
\caption{Evolution of a spherical pulsar with initial angle $\alpha_0
  = 60$  degrees and angular frequency $\Omega_0$. Time on the upper
  horizontal axis is measured in units of characteristic pulsar spin down
  time $\tau = I c^3/\mu^2\Omega_0^2$. \textbf{[Left panel:]} The evolution of angular velocity of
  the neutron star. Vacuum pulsars evolve toward a constant
  rotational frequency, 
  $\Omega_0 \cos \alpha_0$, while MHD pulsars follow power-law
  dependence, $\Omega\propto t^{-1/2}$ for $t\gg\tau$. \textbf{[Right
    panel:]} The evolution of the obliquity angle. In the vacuum case, the
  evolution is exponentially fast and leads to a very sharp decrease
  in the inclination angle. In the MHD case, the evolution is a power-law
  in time, $\alpha \propto t^{-1/2}$.}
\label{fig:spherical}
\end{figure*}

The torque given by equations (\ref{eqn:Kz}) -- (\ref{eqn:Kx}) forces the
star to change its angular velocity as
\begin{equation}
\K = I \frac{{\rm d} {\boldsymbol \Omega}}{{\rm d} t},
\end{equation}
where $I$ is the stellar moment of inertia. The torque component $K_z$
acts in the direction opposite to $\boldsymbol \Omega$ and reduces the
magnitude of angular velocity without changing its direction. On the
other hand, $K_x$ and $K_y$ act perpendicular to $\boldsymbol \Omega$
and change its direction without changing its magnitude. $K_x$ moves
angular velocity towards the magnetic moment $\boldsymbol \mu$, so it
changes the inclination angle $\alpha$. The anomalous torque
component, $K_y$, causes the star to precess on the timescale of
\begin{equation}
T_{\rm anom} =2\pi\tau_{\rm anom} = \frac{2\pi I R c^2}{\mu^2
  \Omega k_3} = \frac{2\pi\xi}{R c \dot P k_3},
\label{eq:tauanom}
\end{equation}
where the prefactor 
\begin{equation}
\xi = k_0 + k_1 \sin^2 \alpha 
\label{eq:xi}
\end{equation}
reflects the inclination-dependence of
pulsar radiation losses (see eq.~\ref{eqn:Kz}). As we will see below, the anomalous torque
does not affect the long-term evolution of spherical stars and has
only a limited effect on the evolution of non-spherical stars.

A spherical star evolves according to,
\begin{align}
\label{eqn:OmegaDot}
I \frac{{\rm d} \Omega}{{\rm d} t} &= K_z,\\
I \Omega \frac{{\rm d} \alpha}{{\rm d} t} &= - K_x.
\label{eqn:AlphaDot}
\end{align}
The solution of pulsar evolution equations could be simplified, because the
system of equations (\ref{eqn:Kz})--(\ref{eqn:Kx}) and
\eqref{eqn:OmegaDot}--(\ref{eqn:AlphaDot}) has an integral
\begin{equation}
\Omega \left( \frac{\cos^{k_0 + k_1}\alpha}{\sin^{k_0} \alpha}\right)^{1/k_2} = \Omega_0 \left( \frac{\cos^{k_0+k_1}\alpha_0}{\sin^{k_0} \alpha_0}\right)^{1/k_2},
\end{equation}
which in the case of vacuum losses ($k_0=0$, $k_1=k_2=2/3$, see
Tab.~\ref{table:models}) reduces to
\begin{equation}
\Omega \cos \alpha = \Omega_0 \cos \alpha_0, \label{eq:vacuum_integral}
\end{equation}
and in the MHD case
($k_0=k_1=k_2=1$) to
\begin{equation}
\Omega \frac{\cos^2\alpha}{\sin\alpha} = \Omega_0 \frac{\cos^2\alpha_0}{\sin\alpha_0}.
\label{eq:MHD_integral}
\end{equation}
Most of the works on the pulsar evolution assume magnetodipole
(vacuum) radiation and ignore the evolution of the obliquity angle,
$\alpha$ \citep{2012Ap&SS.341..457P,2013MNRAS.432..967I}. However, as one can show from equations (\ref{eqn:OmegaDot})--(\ref{eqn:AlphaDot}), the inclination angle evolves on the same
timescale,
\begin{equation}
\tau = \frac{I \Omega}{K_{\rm aligned}} = \frac{I c^3}{\mu^2\Omega^2} = \frac{P\xi}{\dot P},
\label{eq:tau0}
\end{equation}
as the angular velocity, where $\xi$ is given by
eq.~\eqref{eq:xi}. Here $P$ and $\dot P$ are the pulsar period and
period time derivative, respectively.  

As pulsars spin down, the obliquity angle evolves and this affects the
spindown. Therefore, it is crucial to account for the change in
$\alpha$. For
instance, if we do not account for the alignment effect in the vacuum
case, i.e.\ postulate $\alpha = $~constant, a pulsar starting out with
$\Omega = \Omega_0$ and $\alpha = \alpha_0$ evolves towards
$\Omega_{\rm final} = 0$. However, the correct solution of
eqs.~(\ref{eqn:OmegaDot})--(\ref{eqn:AlphaDot}) with torques given by
eqs.~(\ref{eqn:Kz})--(\ref{eqn:Kx}) yields
$\Omega_{\rm final} = \Omega_0 \cos \alpha_0$, which is comparable to
$\Omega_0$ for representative $\alpha_0 \sim 60^\circ$.

Though both vacuum and MHD models evolve toward the aligned configuration, they approach it through quite different paths.
The inclination angle of a vacuum pulsar evolves according to
\begin{equation}
\label{eq:alpha_vac}
\sin\alpha = \sin \alpha_0 \exp\left(-t/\tau^{\rm VAC}_{\rm align} \right),
\end{equation}
where $\tau_{\rm align}^{\rm VAC} = 1.5 \tau \cos^{-2} \alpha_0$,
and $\tau$ is pulsar spindown timescale given by
eq.~\eqref{eq:tau0}. Thus, the vacuum pulsar evolves to the aligned
configuration exponentially fast, without a significant slowdown of its rotation. For an MHD pulsar, we have the following implicit solution,
\begin{equation}
\label{eq:alpha_mhd}
\frac{1}{2 \sin^2 \alpha} + \log(\sin \alpha) = \frac{t}{\tau_{\rm align}^{\rm MHD}} + \frac{1}{2 \sin^2 \alpha_0} + \log(\sin \alpha_0),
\end{equation}
where $\tau_{\rm align}^{\rm MHD} = \tau \sin^2\alpha_0/\cos^4\alpha_0$, which asymptotes at late times to a power-law, $\alpha \propto t^{-1/2}$ \citep{ptl14}.

Figure \ref{fig:spherical} illustrates the evolution under vacuum and
MHD torques of a spherical pulsar with an initial angle of
$\alpha_0 = 60$ degrees and Crab-like parameters, $P = 0.033$~s,
$B = 3.78\times 10^{12}\ {\rm G}$, $\tau = 7.4\times10^3$~years. One can see that
for a vacuum pulsar the inclination angle becomes exponentially small just
after $t \ge \tau^{\rm VAC}_{\rm align}$ which equals to $6\tau$ for
Crab's parameters. On the other hand, the obliquity angle in the MHD case
changes only by a factor of few for the same parameters. Note also that MHD pulsars lose angular momentum more efficiently than vacuum
ones. Instead of pulsar rotation asymptoting to $\Omega_{\rm final}=\Omega_0
\cos\alpha_0$ as for vacuum pulsars, MHD pulsars lose a substantial amount of angular momentum
during its lifetime, and $\Omega$ continuously decreases,
asymptotically approaching zero: $\Omega \propto
t^{-1/2}$. Characteristic timescale of pulsar spindown is
around $10^4{-}10^5$ years (i.e. the timescale over which $\Omega$ changes by a factor of 2).

\section{Evolution of non-spherical stars}
\label{sect:precession}

\begin{figure}
\centering
\includegraphics[scale=0.5]{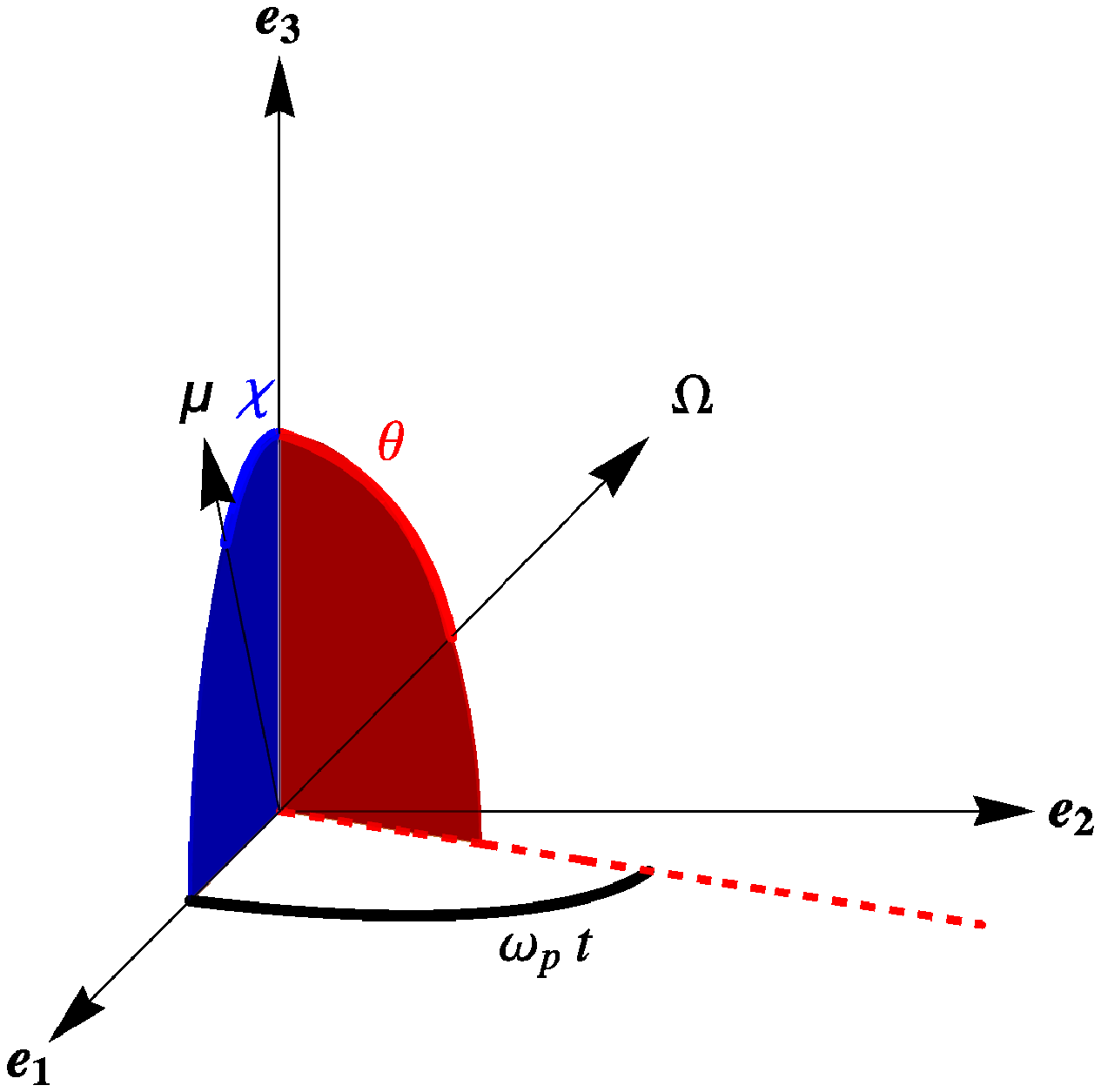}
\caption{
 In this work, we use the following coordinate system: $\boldsymbol e_1$, $\boldsymbol e_2$,
$\boldsymbol e_3$ denote the principal axes of the star, which is
characterised by a stellar magnetic moment $\boldsymbol\mu$ and
angular frequency $\boldsymbol\Omega$. Pulsar geometry is defined by
several angles: $\chi$ between $\boldsymbol\mu$ and
  $\boldsymbol e_3$, $\theta$ between $\boldsymbol\Omega$ and
  $\boldsymbol e_3$, and $\alpha$ (not shown) between
    $\boldsymbol\mu$ and $\boldsymbol\Omega$. During free precession, $\boldsymbol\Omega$
  rotates around $\boldsymbol e_3$ and the angle between the
  $\boldsymbol e_1$ and the projection of $\boldsymbol \Omega$ on
  $\boldsymbol e_1$-$\boldsymbol e_2$ plane increases at a constant
  rate $\omega_p$.}
\label{fig:pulsar}
\end{figure}

\begin{figure*}
\centering
\includegraphics[width=0.5\textwidth]{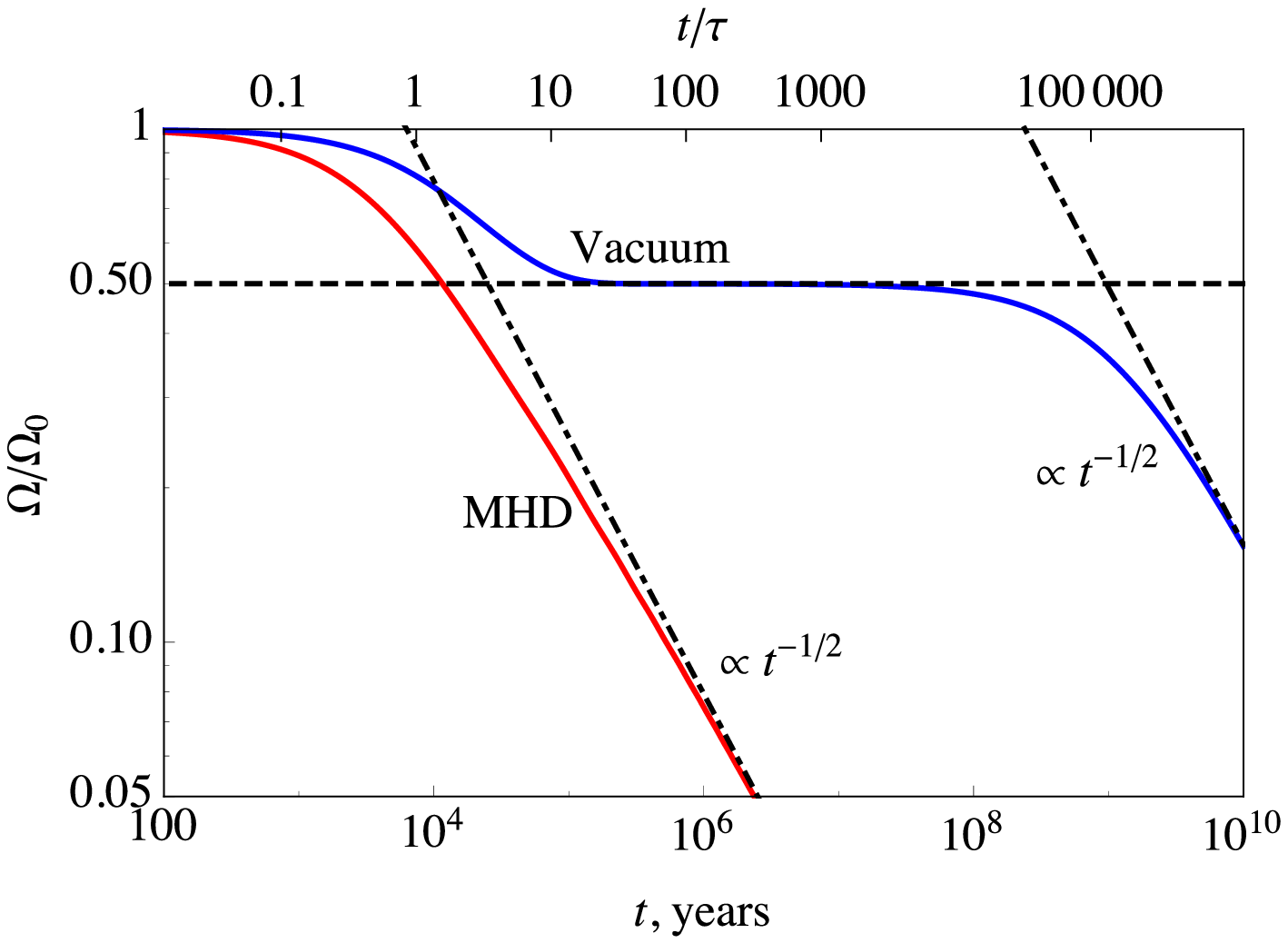}~~~\includegraphics[width=0.5\textwidth]{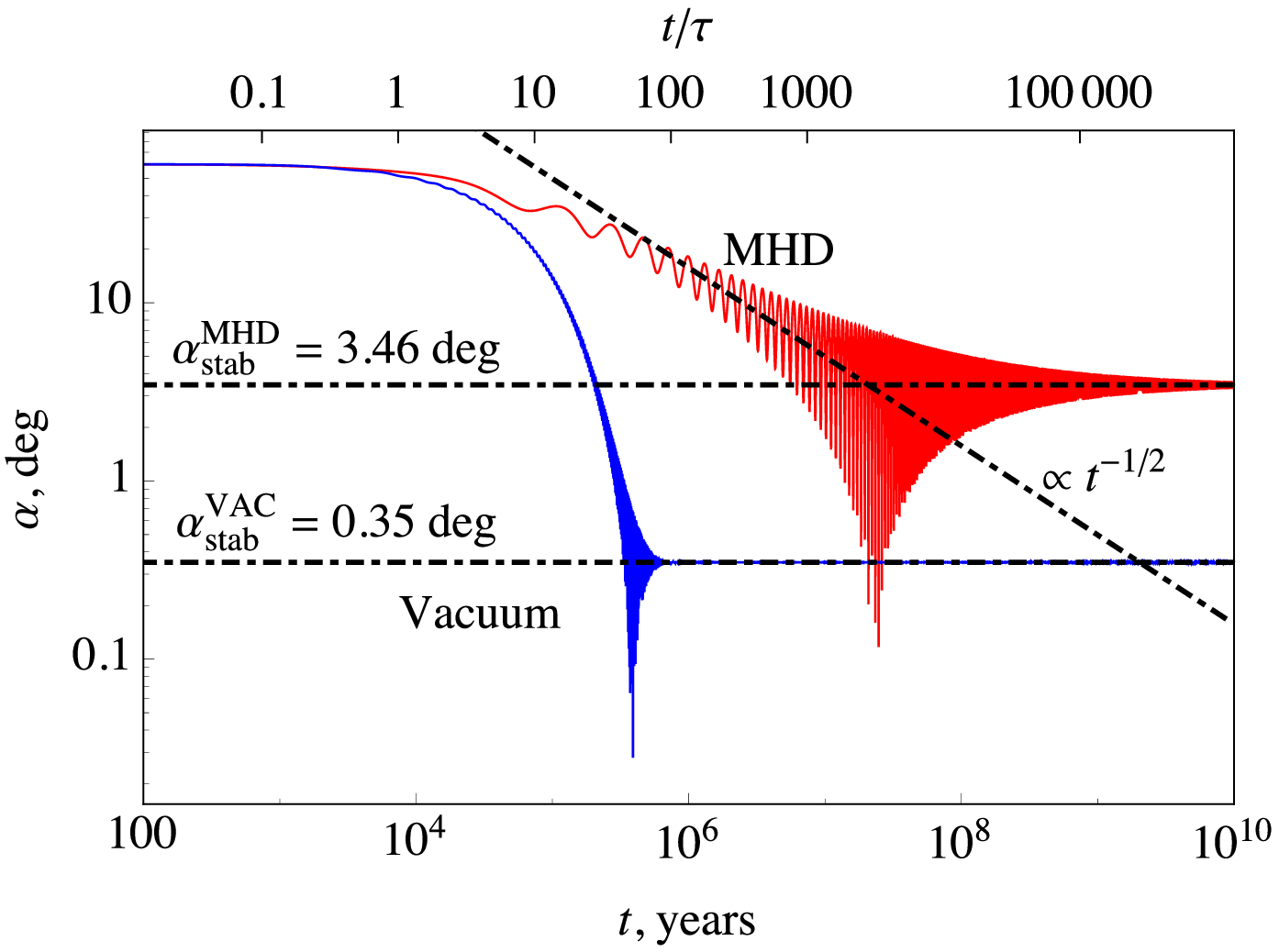}
\caption{The time evolution of a non-spherical pulsar with the initial inclination
  angle $\alpha_0 = 60$ deg, angles $\theta_0 = 60$ deg and
  $\chi_0 = 1$ deg (see Fig.~\ref{fig:pulsar} for the angle definitions), 
  ellipticity $\varepsilon = 1.66 \times 10^{-13}$, and angular
  frequency $\Omega_0$. \textbf{[Left panel:]} The angular velocity
  $\Omega$ vs time. Despite the star is non-spherical, 
  $\Omega$ evolves in a very similar way to the spherical case. \textbf{[Right
    panel:]} The inclination angle $\alpha$ vs time. The stellar
  non-sphericity causes $\alpha$ to oscillate at late time, $t\gtrsim10\tau$ in
  both vacuum and MHD cases. Despite that, in a time-average sense the
  inclination angle undergoes a secular decrease.
Eventually, the inclination angle stabilises. The stabilisation in MHD
pulsars occurs at a much later time than in vacuum pulsars.
}
\label{fig:nonspherical}
\end{figure*}

The rotation of pulsars steadily slows down,
and the radio pulse profile is mostly stable in time. The deviation from smooth
spindown is often less than one part in $10^{11}$, but at this accuracy pulsars
experience rather random irregularities, which are referred to as \emph{timing noise}.

\citet{2010Sci...329..408L} presented timing residuals (deviations
from steady spindown) of 17 pulsars that showed a quasi-periodic
behaviour. The characteristic period of these variations is a few
hundred days, hence prolonged observations are needed to investigate
this behaviour. Interestingly, a lot of pulsars with timing noise
exhibit variations in the pulse profile that are correlated with
variations in the period. This makes a change in geometry (e.g.,
precession-induced changes in
the inclination angle) a possible explanation of this
phenomenon. Below we focus on the two cases of pulsar B1828-11 and the
Crab pulsar.  In this section, we set $R = 10^6$~cm and consider the
case of a plasma-filled magnetosphere, i.e., set
$k_0 = k_1 = k_2 = 1$.
 
Previously, we discussed a spherical star with 
an isotropic tensor of inertia. Generally, a rigid star has a
triaxial inertia tensor that is diagonalised in the basis of its
principal axes.  Let us denote the pulsar principal axes as
$\boldsymbol e_1$, $\boldsymbol e_2$, $\boldsymbol e_3$ and
corresponding moments of inertia as $I_1 = I$,
$I_2 = I(1+\varepsilon_{12})$ and $I_3 = I (1+\varepsilon_{13})$, respectively.
Here, $\varepsilon_{12}$ and $\varepsilon_{13}$ characterise the
degree of stellar non-sphericity.

We will start with the case of a \emph{biaxial} star, i.e., we
will set $\varepsilon_{12} =0$,
$\varepsilon_{13} = \varepsilon \ne 0$. Without the loss of generality, we
choose $\boldsymbol e_1$ to lie in the
$\boldsymbol e_3{-}\boldsymbol \mu$ plane. As seen in
Fig.~\ref{fig:pulsar}, the problem geometry is now set by three
angles: angle $\chi$ between $\boldsymbol \mu$ and $\boldsymbol e_3$,
angle $\theta$ between $\boldsymbol \Omega$ and $\boldsymbol e_3$ and
angle $\alpha$ between $\boldsymbol \Omega$ and $\boldsymbol
\mu$. Note that the
values of these angles are not completely arbitrary but are constrained by
$|\theta - \chi| \le \alpha \le |\theta + \chi|$.  Here and below,
indices \hbox{($1$, $2$, $3$)} refer to the coordinate system set by the principal
axes and \hbox{($x$, $y$, $z$)} to the coordinate system described in
Section~\ref{sect:vacuum}.

For a non-spherical pulsar, instead of
equations (\ref{eqn:OmegaDot})--(\ref{eqn:AlphaDot}) we need to solve
Euler's equations of motion of a rigid body:
\begin{equation}
\dot L_i = \epsilon_{ijk} L_j \Omega_k + K_i,
\label{eqn:euler}
\end{equation}
where summation over repeated indices is implied, $K_i$ and $\Omega_i$ are the components of $\boldsymbol K$ and
$\boldsymbol \Omega$ in the ($\boldsymbol e_1$, $\boldsymbol e_2$,
$\boldsymbol e_3$) coordinate system and $L_i = I_i \Omega_i$.

To compute $K_i$, we need to make a coordinate transformation from
($\boldsymbol x$, $\boldsymbol y$, $\boldsymbol z$) to ($\boldsymbol e_1$, $\boldsymbol e_2$,
$\boldsymbol e_3$). This transformation could be found
in \citet{Melatos}, and we do not give it here. Rewriting
equation (\ref{eqn:euler}) in the new coordinates, we obtain:
\begin{equation}
 \dot\Omega_{1} + \varepsilon \Omega_{2} \Omega_3 = K_{1}/I,
\label{eqn:1}
\end{equation}
\begin{equation}
\label{eqn:2}
\dot\Omega_{2} - \varepsilon \Omega_{3} \Omega_1 = K_{2}/I,
\end{equation}
\begin{equation}
(1+\varepsilon)\dot\Omega_{3} = K_{3}/I.
\label{eqn:3}
\end{equation}

We solve these equations numerically using the three angles
$\theta_0$, $\chi_0$ and $\alpha_0$ and the initial period $P_0$ as
the initial conditions. We adopt the following fiducial parameters for
a neutron star: mass $M = 1.4 M_{\sun}$, radius $R = 10$~km, dipole
moment $\mu = 0.5B R^3$, and $B =10^{12}$~G. 

To see how a small stellar non-sphericity can affect the evolution of
a neutron star, let us again consider a Crab-like pulsar,
$\alpha_0 = 60^{\circ}$, $P = 0.033$~s, $B_{12} = 3.78$,
$\tau = 7.4\times10^3$~years. Fig. \ref{fig:nonspherical}
shows how this pulsar evolves for $\theta_0 = 60^\circ$,
$\chi_0 = 1^\circ$ and $\varepsilon = 1.66 \times 10^{-13}$. The right
panel in Fig.~\ref{fig:nonspherical} shows that even such a seemingly
small ellipticity causes the inclination angle $\alpha$ to vary with a
substantial amplitude. These variations are due to the free precession
of the star, at the period set by the degree of stellar non-sphericity:
\be 
T_{\rm prec} = 2\pi\tau_{\rm prec} = \frac{P}{\varepsilon},
\label{eq:tauprec}
\ee
which for our choice of parameters is $T_{\rm prec} = 6300~{\rm years}$.
As we will see below, the standard expression \eqref{eq:tauprec} applies only in the limit of
angular frequency pointing along the symmetry axis of the neutron
star, $\boldsymbol e_3$, i.e., for $\cos\theta_0\approx1$  [see
equation~\eqref{eq:tauprecaxisymm}, which
generalises eq.~\eqref{eq:tauprec} to other values of $\theta_0$].

In vacuum, the rotation of a \emph{spherical} star asymptotes to a
non-zero value: this is because pulsars quickly evolve toward the
alignment $\alpha = 0$ and their spindown torque vanishes. In contrast,
the left panel in Fig.~\ref{fig:nonspherical} shows that
\emph{non-spherical} vacuum pulsar rotation slows down all the way to
zero.  This is because non-sphericity effects cause the inclination
angle to \emph{stabilise} at a finite value, e.g., at
$\alpha_{\rm stab}^{\rm VAC} = 0.35$~deg for our choice of pulsar
parameters, which prevents the spindown torque from vanishing (see eq.~\ref{eq:Kzvac}).
Similarly, for MHD pulsars, stellar non-sphericity leads to the
inclination angle stabilisation at a finite value, e.g., at
$\alpha^{\rm MHD}_{\rm stab}=3.46^{\circ}$ for our choice of pulsar parameters,
as seen in the right panel of Fig. \ref{fig:nonspherical}. The value
of the stabilisation angle depends on the strength of anomalous torque
and the $k_3$ factor (see Tab.~\ref{table:models}) and can
substantially differ between vacuum and MHD models; a study of this
dependence is beyond the scope of this work.

After stabilisation, the evolution of the angular velocity of the neutron
star can be approximated by
\be
\label{eq:stab}
\langle \dot \Omega \rangle = - \frac{\langle \Omega \rangle^3
  \mu^2}{I c^3} (k_0 + k_1 \sin^2 \alpha_{\rm stab}),
\ee
where brackets $\langle\dots\rangle$ denote the averaging over a precession period. Equation \eqref{eq:stab} has a solution
\be
\label{eq:sol_stab}
\frac{1}{\langle \Omega\rangle^2} - \frac{1}{\Omega_{\rm stab}^2}=
\frac{t}{\tau_{\rm evol}},
\ee
where $\tau_{\rm evol}$ is the characteristic timescale of evolution and
$\Omega_{\rm stab}$ is the value of angular frequency at stabilisation. The
timescale $\tau_{\rm evol}$ is determined by the stabilisation angle and is different in
vacuum and MHD cases:
\begin{align}
\label{eq:tau_vac}
\tau_{\rm evol}^{\rm VAC} &= 1.5 \tau \sin^{-2} \alpha_{\rm stab} \gg \tau, \\
\tau_{\rm evol}^{\rm MHD} &= \tau (1+\sin^2 \alpha_{\rm stab})^{-1} \approx \tau.
\end{align}
Equation (\ref{eq:sol_stab}) implies a plateau in
$\langle \Omega \rangle$ at early times post-stabilisation (see the
left panel of Fig. \ref{fig:nonspherical}) and a power-law evolution
at late times, $t\gtrsim\tau_{\rm evol}$. For the case shown in Figure
\ref{fig:nonspherical} the analytical estimate \eqref{eq:tau_vac}
gives $\tau_{\rm evol}^{\rm VAC} = 5.21 \times 10^8$ years, in
agreement with the end of plateau.

Figure~\ref{fig:nonspherical} shows that until the inclination angle
stabilises, the solution for a non-spherical pulsar is similar to the
solution for a spherically-symmetric neutron star, except for
small-amplitude short timescale oscillations super-imposed on top of
the secular alignment trend. This allows us make use of
eqs.~(\ref{eq:alpha_vac}) and (\ref{eq:alpha_mhd}) to obtain a simple
estimate of the time at which the stabilisation sets~in:
\begin{align}
t_{\rm stab}^{\rm VAC} 
&=  1.5 \tau \frac{\log(\sin\alpha_0/\sin\alpha_{\rm stab})}{\cos^2 \alpha_0}
\sim {\rm few} \times \tau,\label{eq:tstabvac}\\
t_{\rm stab}^{\rm MHD }&\approx  \frac{\tau}{\alpha_{\rm
  stab}^2}.
\end{align}
Here we assumed the stabilisation angle to be small and neglected the
logarithmic factor in eq.~\eqref{eq:tstabvac}. Importantly, the stabilisation
time for MHD pulsars is much larger than for vacuum ones:
\be
\frac{t_{\rm stab}^{\rm MHD }}{t_{\rm stab}^{\rm VAC}} \sim
\frac{1}{{\rm few}\times\alpha_{\rm
  stab}^2} \gg 1,
\ee
but even for the vacuum pulsar this is a large timescale: for the
initial conditions used in Fig.~\ref{fig:nonspherical}, $t_{\rm stab}^{\rm VAC}
\sim 10^5$ years. For MHD pulsars it is $\gtrsim10^8$
years (for $\alpha_{\rm stab} \lesssim 0.03$). We thus expect that pulsars cross the death line before
their inclination angle has a chance to reach a small value and become
stabilised.

\subsection{Neutron Star Ellipticities}
\label{sect:ellipticities}

Due to the uncertainties in the neutron star equation of state
\citep{1997JPhG...23.2013G}, it is hard to
estimate how much the pulsar crust is deformed and therefore to
theoretically compute the value of ellipticity. \citet{2009PhRvL.102s1102H} performed
multi-million ion molecular dynamics simulations showing that the
neutron star crust can support ellipticities up to
$\epsilon_{\rm max} = 4 \times 10^{-6}$. This ellipticity corresponds
to a precession time of a few days for a one-second pulsar (see eq.~\ref{eq:tauprec}).

There are two main contributors to the neutron star ellipticity: fast
rotation and strong magnetic fields in the interior. The non-sphericity caused by
rotation can be estimated as the ratio between its rotational and
gravitational energies:
\begin{equation}
\label{eqn:ell_rot}
\varepsilon_{\rm rot} = \frac{E_{\rm rot}}{E_{\rm grav}} 
\approx 7\times 10^{-8} P_1^{-2} R_6^3M_{1.4}^{-1},
\end{equation}
where $R = R_6\times 10^6$~cm, $P = P_1\times 1$~s, and $M=M_{1.4}
\times 1.4M_\odot
$. This deformation is pointed along the rotational axis. 

However, not the entirety of this ellipticity participates
in precession.  The neutron star supports hydrostatic
stresses only after the crystallisation of its crust. The part of the
ellipticity that participates in precession is much smaller
than the value given by eq.~(\ref{eqn:ell_rot}) and can be written as
\citep[see, e.g.,][]{1975regd.book.....M}:
\begin{equation}
\label{eqn:ell_cr}
\varepsilon_{\rm cr} = \frac{\tilde \mu}{1 + \tilde \mu} \varepsilon_{\rm rot} = 2 \times 10^{-11} \mu_{30} P^{-2} R_6^7 M_{1.4}^{-3},
\end{equation}
where $\tilde \mu = 19\mu^{\rm cr}/2\rho g R$, where $g$ is the
surface gravity, $\rho$ is the density and $\mu^{\rm cr}$ is shear
modulus of the crust, whose fiducial value is $10^{30}\ {\rm dyn\; cm^{-2}}$.

The impact of the magnetic field can be calculated in a similar way:
\begin{equation}
\label{eqn:ell_mag}
\varepsilon_{\rm mag} = \frac{E_{\rm mag}}{E_{\rm grav}} \approx
10^{-12}B_{12}^{2} R_6^4 M_{1.4}^{-2},
\end{equation}
where $B = B_{12} \times 10^{12}\ {\rm G}$. This deformation is pointed along
the magnetic moment. We discuss the possible origin of stellar
non-sphericity in Section \ref{sect:conclusions}.

\subsection{Pulsar B1828-11}
\label{sec:b1828-11}

Pulsar B1828-11 was the first to show highly periodic long-term
variations in timing residuals that are correlated with variations in
the pulse profile. The spectrum of timing residuals
consists of three harmonics with periods 1000, 500 and 250 days. If
the period of free precession $P/\varepsilon$ (see eq.~\ref{eq:tauprec})  equals the largest
period in spectrum, ellipticity equals $\varepsilon \approx 4.69 \times 10^{-9}$.

We will use the following timing parameters for pulsar B1828-11: $P^0 = 0.405$~s, $\dot
P_{-15}^0 = 60$. The values with the superscript `0' denote the best fit
parameters, and the values without this superscript denote the
observations. The characteristic timescales of the pulsar
evolution are:
\begin{align}
\tau &= 6.75 \times 10^{12}\xi~{\rm s}\\
\tau_{\rm anom} &= 3.49 \times 10^{9} \xi k_3^{-1}~{\rm s},
\end{align}
where $\tau$ is spindown timescale given by eq.~\eqref{eq:tau0} and $\xi = 1 +
\sin^2 \alpha$ is a number between~$1$~and~$2$ that reflects the
dependence of MHD pulsar spindown on the inclination angle (see eq.~\ref{eqn:Kz}).

\begin{figure*}
\centering
\includegraphics[scale=0.7]{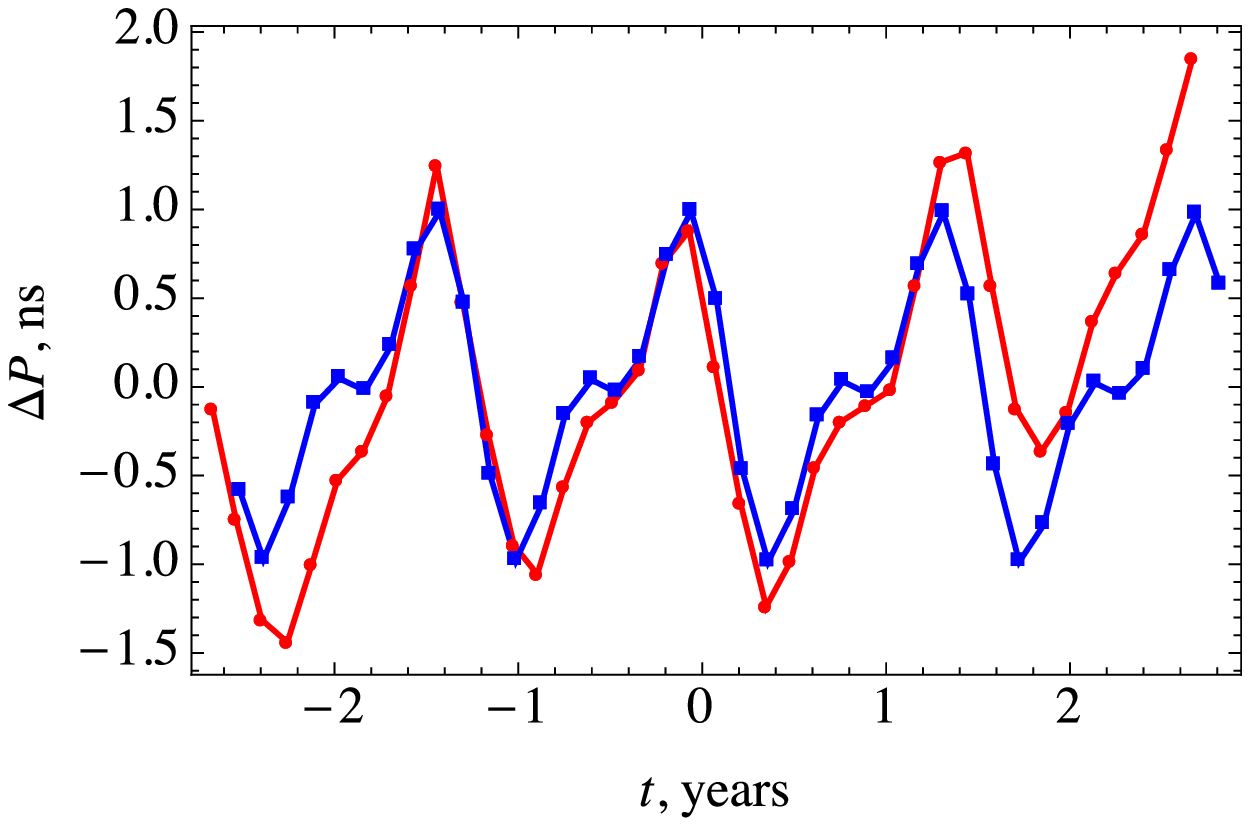}~~~\includegraphics[scale=0.7]{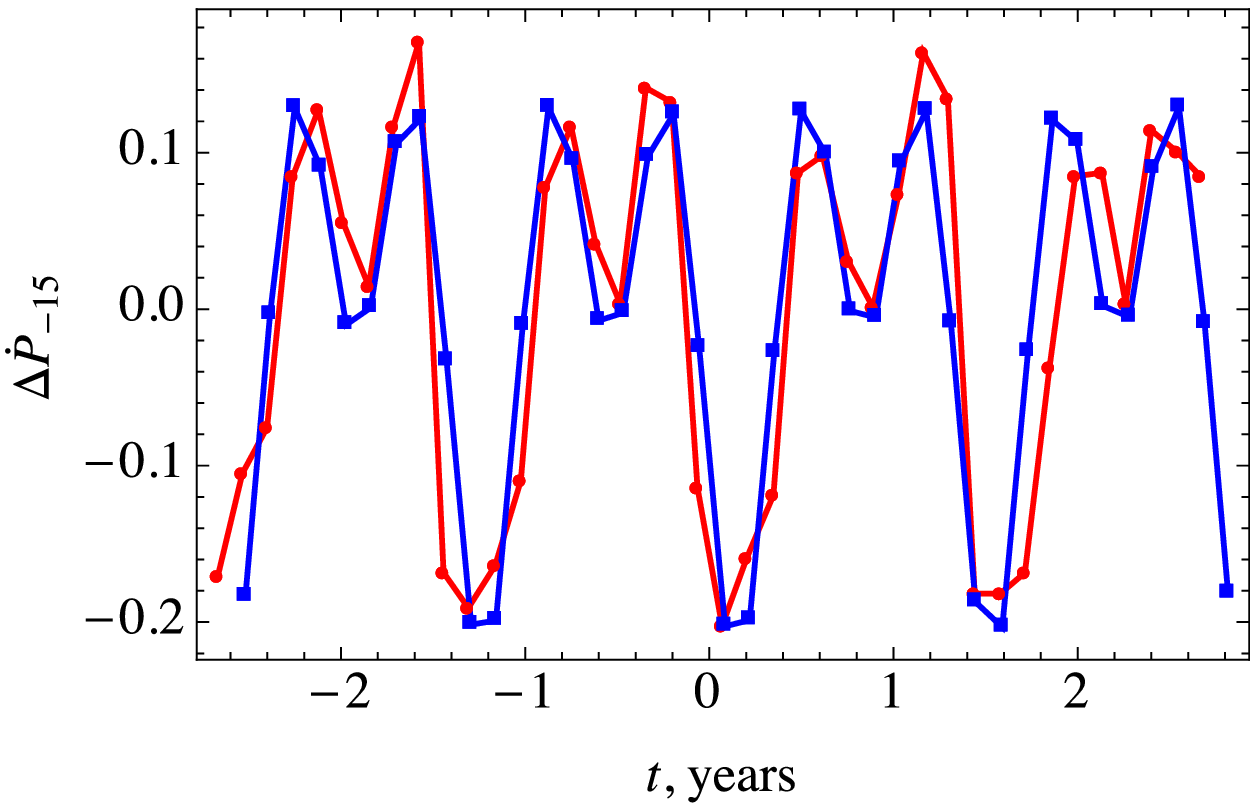}
\caption{Numerical solution of Euler's equations (blue squares) giving
  our best-fit for PSR  B1828-11, $\chi = 89$ deg, $\theta = 6$ deg,
  and the observed
  residuals (red dots). \textbf{[Left panel:]}
  Period residual. The numerical solution is in good agreement with
  observations, reproducing both the characteristic period and the amplitude. \textbf{[Right panel:]} Period derivative
  residual. Our numerical solution reproduces the two-bump structure well.}
\label{fig:b1828}
\end{figure*}

This pulsar also has a well-measured second period derivative
$\ddot P^0 = -1.70 \times 10^{-25} s^{-1}$ and the changes in the
pulse profile are correlated with the changes in $P$ and $\dot P$. As
seen in Fig.~\ref{fig:b1828}, the period timing residual,
$\Delta P = P - P^0-\dot P^0 t - \ddot P^0 t^2/2$, is periodic and has
an amplitude of 1.65 ns, and the residual for the period derivative,
$\Delta \dot P = \dot P - \dot P^0 -\ddot P^0 t$, is also periodic and
has a magnitude of about $0.19\times 10^{-15}$.  In order to compare
the outcome of our models to the observations, i.e., compare the
simulated values of $\Delta P$ and $\Delta \dot P$ to the observations
in Fig.~\ref{fig:b1828}, we will use the same procedure as was used to
determine the observed values of residuals \citep{2013Sci...342..598L}:
every 50 days (indicated by dots in the figure), we average the residual over the
time interval of 100 days.

\citet{2000Natur.406..484S} concluded that the most probable explanation
for these variations is the precession of the neutron star. \hbox{\citet{Link}}
presented a calculation, where
they attempted to find the best-fitting pulsar geometry under the vacuum approximation, that matches
the residuals, $\Delta P$, and the changes in the emission profile. We
now look for a solution under a more realistic assumption of the
plasma-filled magnetosphere.

To reproduce the observational data for pulsar B1828-11, we
set $\tau_{\rm prec} = \pm 0.218$ years.  The timescale can formally be negative
because of the negativity of $\varepsilon$; for the comparison of time
scales, what matters is its absolute value, which corresponds to the period of
variation of about $500$ days -- the leading harmonic in the spectrum -- and
$\varepsilon = \pm 9.44 \times 10^{-9}$. One can also show that for a
biaxial star, the timing residuals do not depend on the sign of the ellipticity.

We numerically solve equations (\ref{eqn:1})--(\ref{eqn:3}) for
different initial values of $\alpha$,
$\theta$ and $\chi$. After that, we find the mean value of the period and its
first two derivatives, and compute the residuals:
\begin{align}
\delta P &= P - \langle P
\rangle - \langle\dot P\rangle t - \langle\ddot P\rangle t^2/2,\\
\delta \dot P &= \dot P - \langle\dot P\rangle - \langle\ddot P\rangle t.
\end{align}

Angles $\theta$, $\chi$ and $\alpha$ are set to their initial values (which
are denoted with the subscript `0') at time $t = 0$ that corresponds to 50,300 MJD.
Euler's equations are solved in $t$ from $-3$ years to $3$ years to
mimic the observed data range. The mean values are calculated over the
entire time range.

\begin{table}
\centering
\caption{\textbf{Pulsar B1828-11:} Best fit parameters obtained by
  fitting the pulsar timing residuals (see Fig.~\ref{fig:b1828}).}
\begin{tabular}{lccc}
\hline
 Case & $\theta$, deg & $\chi$, deg & $\varepsilon_{13}$\\
\hline
\hline
large $\chi$ & 5 & 89 & $9.4 \times 10^{-9}$ \\
small $\chi$ & 89  & 5 &$5.4 \times 10^{-7}$ \\
\hline
\end{tabular}
\label{table:result}\\
\end{table}

Our best fit gives $\chi = 89$
deg and $\theta = 6$ deg and is presented in Fig. \ref{fig:b1828} and
in Table~\ref{table:result}. The
solution implies that the evolution of a neutron star is close to free
precession with small perturbations caused by the magnetospheric
torque. Inclination angle $\alpha$ varies from $\chi-\theta = 83$ deg
to $\chi + \theta = 95$ deg and its initial value simply sets the
initial phase of the evolution. As we will discuss below, a symmetry
exists between the values of $\theta$ and $\chi$. This symmetry implies the
existence of a ``mirror'' solution with the values of $\theta$ and
$\chi$ swapped (see Table~\ref{table:result}). To distinguish between
these two solutions, we refer to them as the ``large $\chi$'' and
``small $\chi$'' solutions. For brevity, we do not
show the small $\chi$ solution in Figure~\ref{fig:b1828}.

In order to reproduce the magnitude of the observed residuals, in our
numerical solution we need to choose the angle $\theta$ to be small and
the angle $\chi$ large, or vice versa: as we discuss below,
this is required for the residuals to have such a small amplitude as
observed, as well as the two-peak structure of the $\dot P$ residual. 

\subsection{Analytical solution for B1828-11}
\label{sect:analytics}

Our numerical solution for pulsar B1828-11 implies that the motion of
a slightly non-spherical neutron star is essentially a free precession
with a small perturbation caused by the magnetospheric
torques.\footnote{So long as we seek a solution for the time interval
  much smaller than the alignment time.}

\subsubsection{Free precession}
For freely precessing body, Euler's equations of motion take the
following form:
\begin{equation}
\dot{\boldsymbol L} + {\boldsymbol \Omega} \times \boldsymbol L = 0.
\end{equation}

These equations can be solved analytically in terms of
Jacobian elliptic functions $\rm cn,~sn~and~dn$ \citep{LLM}:
\begin{align}
L_1/I \Omega_0 &= \sin\theta_0 \; {\rm cn} (\omega_{\rm p} t, k \tan^2 \theta_0),\\
L_2 /I \Omega_0 &= \sin\theta_0(1+k)^{1/2} \; {\rm sn} (\omega_{\rm p} t, k \tan^2 \theta_0),\\
L_3 /I \Omega_0 &= \cos\theta_0 \; {\rm dn} (\omega_{\rm p} t, k \tan^2 \theta_0),
\end{align}
where
\begin{equation}
k = \frac{I_3 (I_2 - I_1)}{I_1 (I_3 - I_2)} 
= \frac{\varepsilon_{12} (1+\varepsilon_{13})}{\varepsilon_{13}-\varepsilon_{12}} 
\approx \frac{\varepsilon_{12}}{\varepsilon_{13}-\varepsilon_{12}}
\label{eq:kellipticity}
\end{equation}
and
\begin{equation}
\omega_{\rm p} = \frac{\varepsilon_{13} L \cos\theta_0}{I_3 (1+k)^{1/2}}
\approx \frac{\varepsilon_{13}\Omega_0 \cos\theta}{(1+k)^{1/2}}.
\label{eq:omegap}
\end{equation}
The precession
period is equal
\begin{equation}
\label{eq:tauPrec}
T_{\rm prec} = \frac{P}{\varepsilon_{13} \cos \theta_0} \frac{2 F(\pi/2,
  k \tan^2\theta_0)}{\pi},
\end{equation}
where $F(\phi, m)$ is the Legendre elliptic integral of the first kind.

In the case of $\varepsilon_{12} = 0$, that is of an axisymmetric star, the
solution takes a much simpler form:
\begin{align}
L_1 /I \Omega_0 &= \sin\theta_0 \cos (\omega_{\rm p} t),\\
L_2 /I \Omega_0 &= \sin\theta_0 \sin (\omega_{\rm p} t),\\
L_3 /I \Omega_0 &= \cos\theta_0, 
\end{align}
where $\omega_{\rm p} = \varepsilon_{13} \Omega_0 \cos \theta_0$ and $\Omega_0$
is the initial rotational frequency. In this case, the precession
period is simply 
\begin{equation}
T_{\rm prec}=\frac{P}{\varepsilon_{13} \cos \theta_0}. \quad {\rm (biaxial\ star)}
\label{eq:tauprecaxisymm}
\end{equation}
Now we have an exact solution $\boldsymbol \Omega_0 (t)$ for free
precession and can use perturbation theory to find the solution
for a nonzero magnetospheric torque.

\subsubsection{Full perturbative solution for a biaxial star}
We find the exact solution for the equation
\begin{equation}
\dot{\boldsymbol L} + {\boldsymbol \Omega} \times \boldsymbol L =
\boldsymbol K \left[\boldsymbol \Omega_0(t)\right].
\end{equation}
in Appendix \ref{sect:appendix}. Here we summarise the most important results.

For an axisymmetric (biaxial) MHD pulsar, the residual in $\dot P$
takes the following form (see Appendix~\ref{sect:twoPeak} for a derivation):
\begin{equation}
\delta \dot P = - \dot P^{\rm obs} f(\theta_0,\chi) [\cos\omega_{\rm
  p}t+g(\theta_0,\chi) \cos 2\omega_{\rm p}t]
\label{eq:timing_residuals}
\end{equation} 
where $\dot P^{\rm obs}$ is the observed value of $\dot P$ and 
\begin{align}
\label{eq:f}
f(\theta,\chi) &=
\frac{\sin2\theta\sin2\chi}{4-2\cos^2\theta\cos^2\chi-\sin^2\theta\sin^2\chi},\\
\label{eq:g}
g(\theta,\chi)&=\tan\theta\tan\chi/4.
\end{align}

The amplitude of the residual is proportional to $\sin 2\theta \sin 2
\chi$, and the residual does not depend on the anomalous torque. 
If we have residuals that are much smaller than $\dot P^{\rm
  obs}$ (which is usually the case), the best-fit values of $\theta$ and $\chi$ are pushed
close either to $0$ or to $90$ deg. As the observed residuals for B1828-11 are much smaller
than $\dot P^{\rm obs}$ and we observe the two-hump structure (and thus have
$g(\theta,\chi)\approx 1$), one of the angles is close to
0, and the other one close to 90 deg.

One can see that expressions (\ref{eq:f}) and (\ref{eq:g}) are
symmetric in respect to the substitution
$\theta \leftrightarrow \chi$.  This means that we cannot distinguish
between the cases of small $\theta$ and small $\chi$, which, as we
mentioned previously,
we refer to as the cases of large $\chi$ and small $\chi$,
respectively. For small $\theta$, the stellar deformation points in
the direction of stellar rotation and it is natural to assume that it
is caused by rotation. If so, it should be on the order of the
analytic expectation, $\varepsilon_{\rm cr}$
(eq.~\ref{eqn:ell_cr}). On the other hand, for small $\chi$ it is
natural to assume that the deformation is caused by the magnetic
field.  If so, we would expect the ellipticity to be on the order of
$\varepsilon_{\rm mag}$ (eq.~\ref{eqn:ell_mag}).

The special case of a biaxial star corresponds to $k=0$ (see
eq.~\ref{eq:kellipticity}). For $k=0$ and a small $\theta$ value, we
have $\cos \theta \sim 1$ and $T_{\rm prec} = P/\varepsilon$ (see
eq.~\ref{eq:tauprecaxisymm}). This implies
$\varepsilon = 9.4 \times 10^{-9}$.  For $k=0$ and small $\chi$, we
find $\cos \theta \sim (\pi/2 - \theta)$ and
$T_{\rm prec} = P/\varepsilon(\pi/2 - \theta)$. For $\theta =89$ deg,
this implies $\varepsilon = 5.4 \times 10^{-7}$. These results are
summarised in Tab.~\ref{table:result}. We discuss the possible origin
of the ellipticity in Sec.~\ref{sect:conclusions}. Note that for our numerical
solution shown in Fig.~\ref{fig:b1828} we chose a small $\theta$ and
large $\chi$. For brevity, we do not show the opposite case.

\begin{figure}
\centering
\includegraphics[scale=0.62]{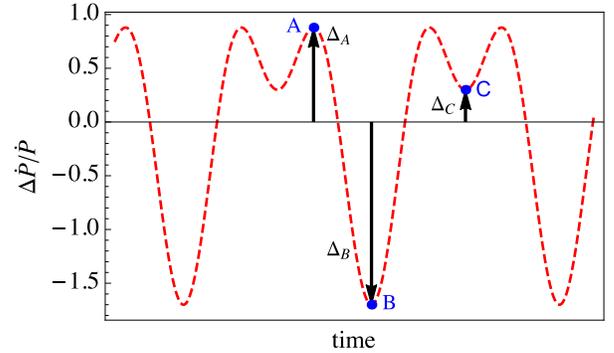}
\caption{Method of obtaining geometrical parameters of precessing neutron stars
  from the observations. One needs to measure the residual values at three extremum
  points ($\Delta_{\rm A}$, $\Delta_{\rm B}$, $\Delta_{\rm C}$) and
  then calculate $f (\chi,\theta)$ and $g(\chi,\theta)$ using
  equations (\ref{eq:A})-(\ref{eq:C}). After that, $\chi$ and $\theta$
  are implicitly set by the values of $f$ and $g$.}
\label{fig:obs}
\end{figure}

\begin{figure*}
\centering
\includegraphics[scale=0.45]{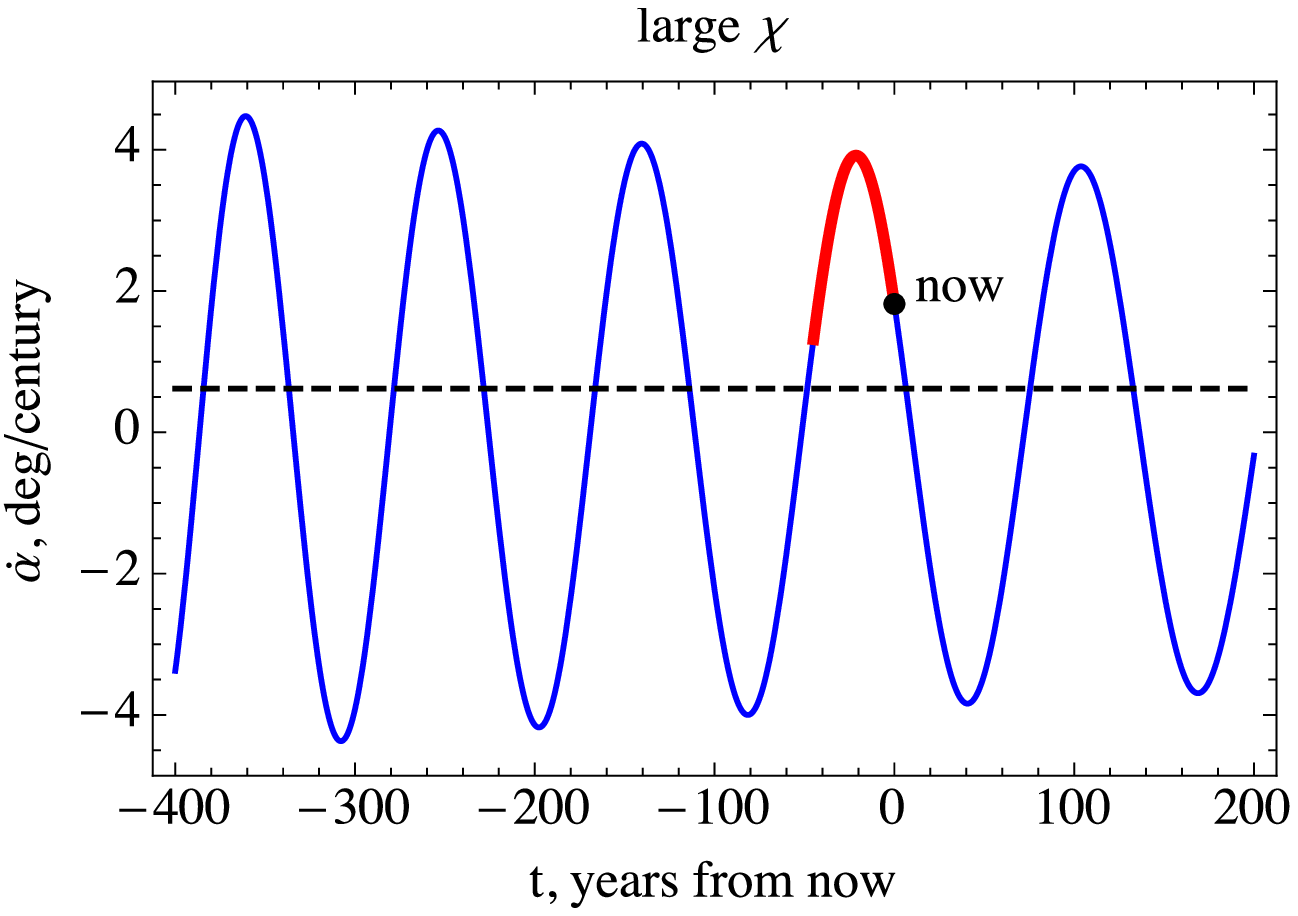}~~\includegraphics[scale=0.45]{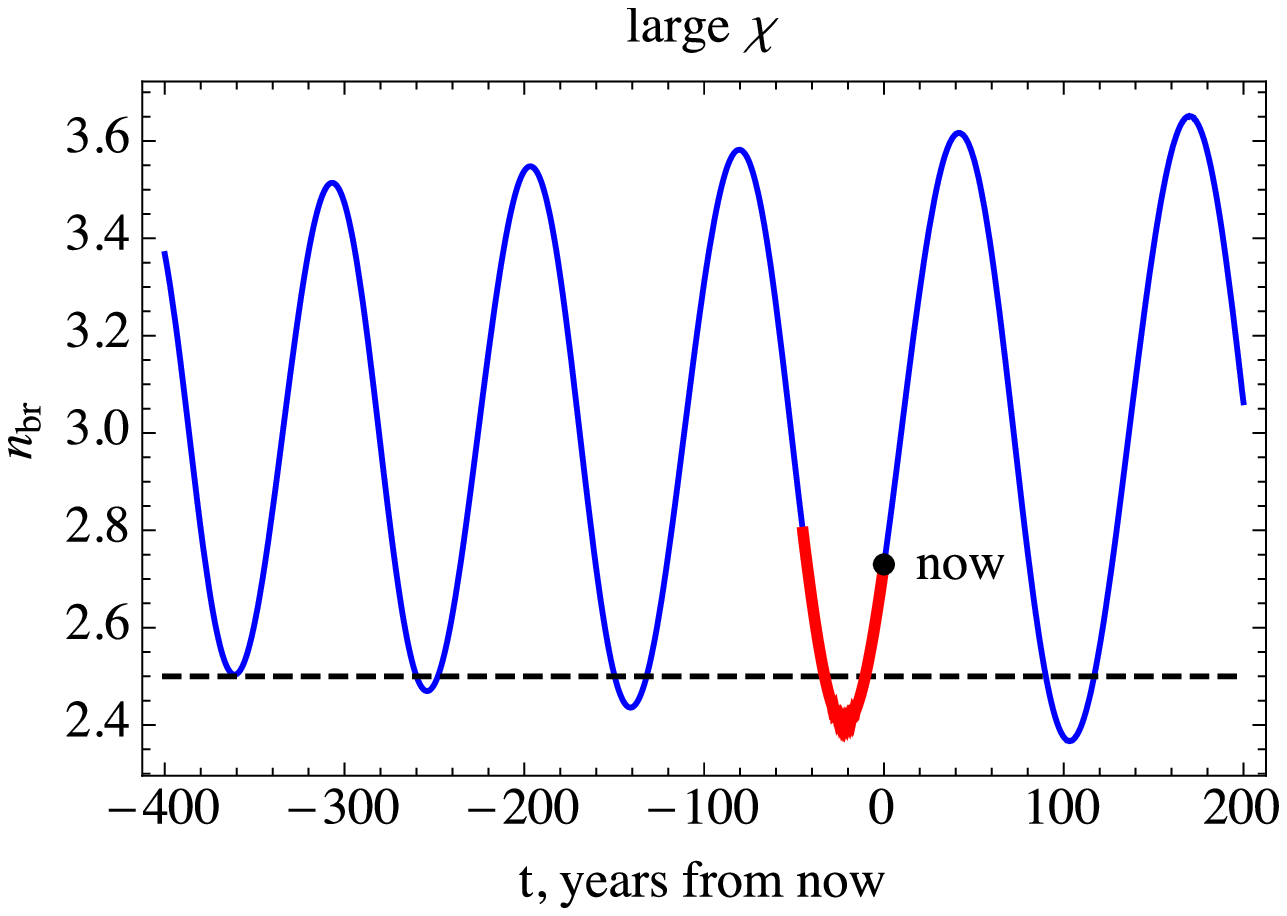}~~\includegraphics[scale=0.45]{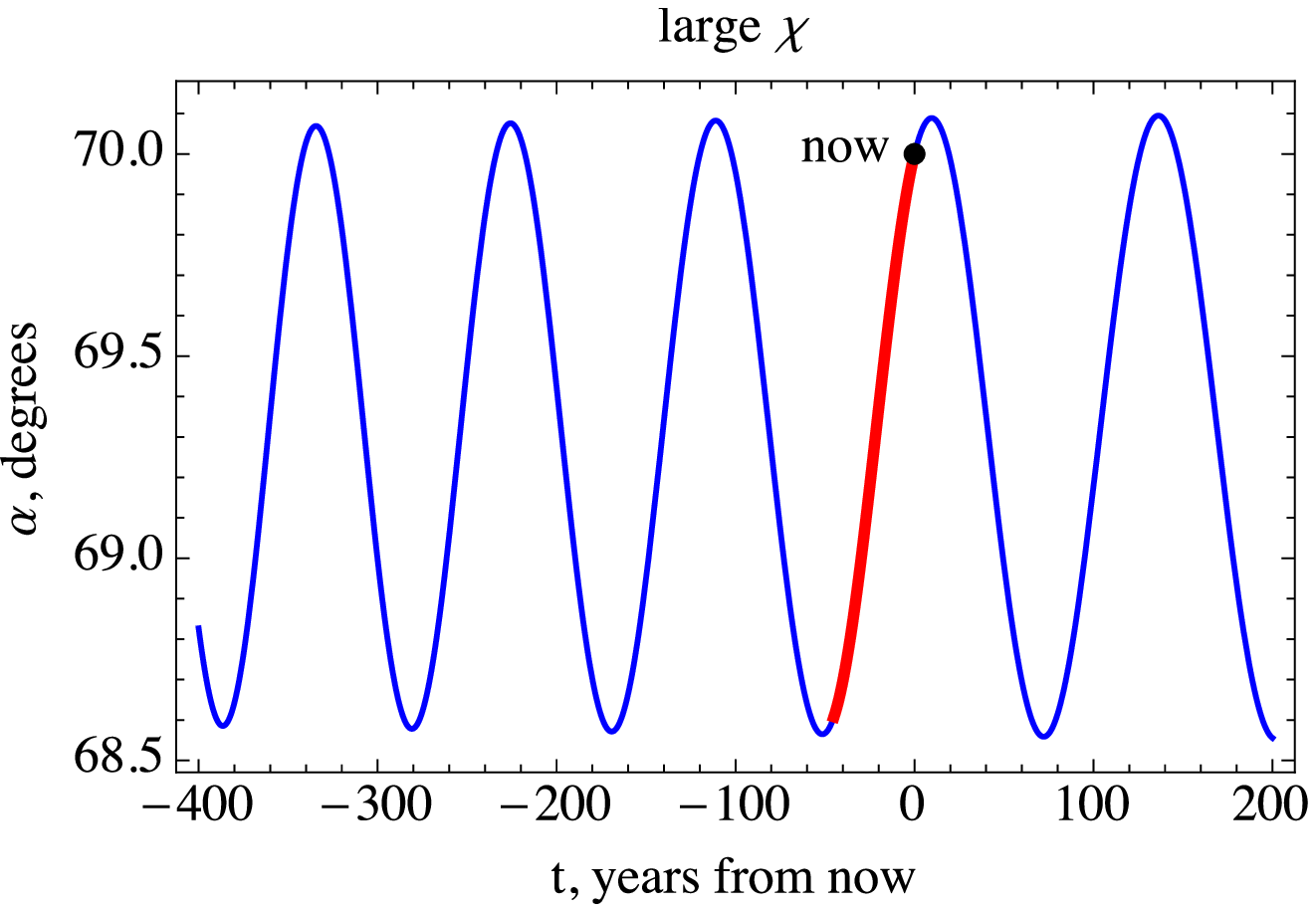}
\caption{Blue line shows our best-fit numerical solution of Euler's equations for the
  Crab pulsar in the \emph{large} $\chi$ case is shown
  (see Fig.~\ref{fig:crab2} for the small $\chi$ case). This red
  line indicates the time range over which timing residuals have
  been observed and the black dot indicates the present time. Black dashed lines show the observationally-inferred values. The pulsar
  parameters at $t=0$ are: $\alpha_0 = 70^\circ$, $\theta_0 = 2^\circ$,
  $\chi_0 = 68.05^\circ$. \textbf{[Left panel:]} The rate of change of
  the inclination angle, $\dot\alpha$. The solution shows an order of magnitude agreement
  with observations. \textbf{[Central panel:]} The braking index
  $n_{\rm br}$ vs time. \textbf{[Right panel:]} The inclination
  angle $\alpha$ vs time. Over the observed time range, $\alpha$ increases almost
  linearly with time, in agreement with observations.}
\label{fig:crab1}
\end{figure*}

Using the analytic model we have developed, we now can estimate
directly from the observations the values of angles $\theta$ and $\chi$, or,
equivalently, the values of the functions $f(\theta,\chi)$ and
$g(\theta,\chi)$ (see eqs.~\ref{eq:f} and \ref{eq:g}). For this, we
only needs to measure the following three extremum points, as shown in
Fig. \ref{fig:obs}: the global maximum $\Delta_{\rm A}$ and minimum
$\Delta_{\rm B}$ as well as
the local minimum $\Delta_{\rm C}$ of the residual,
\begin{eqnarray}
\Delta_{\rm A} & =& f (\theta_0, \chi) \left[g(\theta_0, \chi) + \frac{1}{8 g(\theta_0, \chi)}\right] ,\label{eq:A}\\
\Delta_{\rm B} & =& - f(\theta_0, \chi) [1 + g(\theta_0, \chi)],\\
\Delta_{\rm C} & =& f(\theta_0, \chi) [1 - g(\theta_0, \chi)],\label{eq:C}
\end{eqnarray}
invert these to obtain the values of $f$ and $g$. Note that this
system of equations is over-constrained, and this offers a
self-consistency check. 

\subsubsection{Full perturbative solution for a triaxial star}
If the star is not axisymmetric, the solution has a much more complex
form (but still does not depend of the anomalous torque, see
Appendix~\ref{sect:secondEll} for a derivation):
\begin{equation}
\label{eqn:result}
\delta \dot P = - \dot P^{\rm obs} f(\theta_0,\chi) \mathcal{P}(\theta_0, \chi,t)
\end{equation}
with
\begin{align}
\label{eqn:resultT}
\mathcal{P} &= {\rm cn}(\omega_p
  t)  {\rm dn}(\omega_p
  t) +  {\rm cn}(2\omega_p
  t) \frac{2 g(\theta_0,\chi)}{(1+ {\rm dn}(2\omega_p
  t))} +\\
&+( {\rm dn}(2\omega_p
  t)-1) \left(\frac{g(\theta_0,\chi)}{(1+ {\rm dn}(2\omega_p
  t))} + \frac{ g(\theta_0,\chi)^{-1}}{8(1+ {\rm cn}(2\omega_p
  t))}\right).\nonumber
\end{align}
Here we for simplicity do not write the second argument of Jacobian elliptic functions which always equals $k \tan^2 \theta_0$. 
The second ellipticity does not qualitatively change the behaviour of solution and only influences the precession period (see eq. \ref{eq:tauPrec}).

\subsection{Crab pulsar}
\label{Sec:Crab}

The emission from the Crab pulsar (pulsar B0531+21) was observed
since 1969 in radio, optical, x-ray and gamma-ray wavebands. The radio
pulse profile of this pulsar consists of two components separated by
$145^\circ$ in phase: the main pulse and the interpulse. During high-precision
daily observations performed since year 1991 at Jodrell Bank Observatory,
\citet{2013Sci...342..598L} found a steady increase in separation between the main pulse
and the interpulse, at $0.62^\circ \pm 0.03 ^\circ$ per century.

Because radio and gamma-ray pulses of the Crab pulsar arrive at the
same pulse phase, it is thought that the radio emission from the Crab
pulsar comes from the same magnetospheric region as the gamma-ray
emission. Using this fact, as well as gamma-ray emission models,
\citet{2013Sci...342..598L} concluded that the increase in the separation between the main pulse and
the interpulse indicates an \emph{increase} of the inclination angle at approximately the
same rate. The increase of the inclination angle also naturally
leads to a braking index value less than 3, in agreement with
observations. However, for a spherically symmetric star, MHD models predict a \emph{decrease} of the inclination angle
for a spherical star (see Sect. \ref{sect:sphere}), and a braking index that is always larger than
3. \citet{ptl14} suggested that the observed increase
of the inclination angle could be due to the precession of the
neutron star caused by stellar non-sphericity.

Later, \citet{2015MNRAS.446..857L} revealed the observational data for the Crab pulsar
over a 45-year time range. Although the timing residuals show some
oscillatory behaviour, they are corrupted by a large number of
glitches. This makes it impossible to get an accurate precession model
fit. Nevertheless, between the subsequent glitches Crab pulsar behaves
as if it were precessing.

The observed increase in the inclination angle of the Crab pulsar can
be explained with a precession model. Just as for PSR B1828-11,
the timing residuals residuals in $P$
and $\dot P$ \citep{2015MNRAS.446..857L} imply that either
$\theta\approx0$ and $\chi\approx90$~deg or $\chi\approx0$ and $\theta\approx90$~deg. 

\begin{table*}
\centering
\caption{\textbf{Crab pulsar:} Best fit parameters for large and small
  $\chi$ cases. The second and third frequency derivatives agree with
  observations to within an order of magnitude, as expected given the
  large number of Crab's glitches.   The best-fit values of the inclination angle
  $\alpha$ are agreement with independent estimates
  \citep{2013Sci...342..598L}. Our solution implies a precession
  period of about $100$ years.  Deviations from the near-linear
  increase in time of the inclination angle could
  be observed on shorter timescales, perhaps as short as $\sim 20$ years.}
\begin{tabular}{lccccccc}
\hline
Case & $\theta$, deg & $\chi$, deg & $\alpha$, deg & $T_{\rm prec}$, years & $\varepsilon$ & $\ddot\nu,~10^{-20}$~Hz/s$^2$ & $\dddot \nu,~10^{-30}$~Hz/s$^{3}$\\
\hline
\hline
observations & - & - & - & - & - &  1.12 & -2.73\\
large $\chi$ & 2 & 68.05 & 70 &  126.7 & $8.3 \times 10^{-12}$ & 1.15 & -0.8\\
small $\chi$ & 59.1  & 1  & 60 & 108.5 & $1.11 \times 10^{-11} $ & 1.27 & 0.8\\
\hline
\end{tabular}
\label{table:result_crab}\\
\end{table*}

Figure \ref{fig:crab1} shows our best fit obtained for large $\chi$ case
and $\alpha = 70^\circ$. One can see that for the past 45 years the
inclination of the Crab pulsar has increased almost linearly. On
larger timescale, the inclination angle is oscillating and its
mean value is almost constant. Our fit
produces values of $\dot \nu$, $\ddot \nu$ and $\dddot \nu$ close to the observed values. We
summarise the inferred pulsar geometry
in Table \ref{table:result_crab}. The precession period in
this solution is roughly $T_{\rm prec} = 126$ years, but there is a
strong degeneracy between values of $\theta$ and $T_{\rm
  prec}$. Making precession period $T_{\rm prec}$ larger (which is
allowed by the data) and the wobble angle
smaller, one could obtain sufficiently good fits.  

\begin{figure*}
\centering
\includegraphics[scale=0.45]{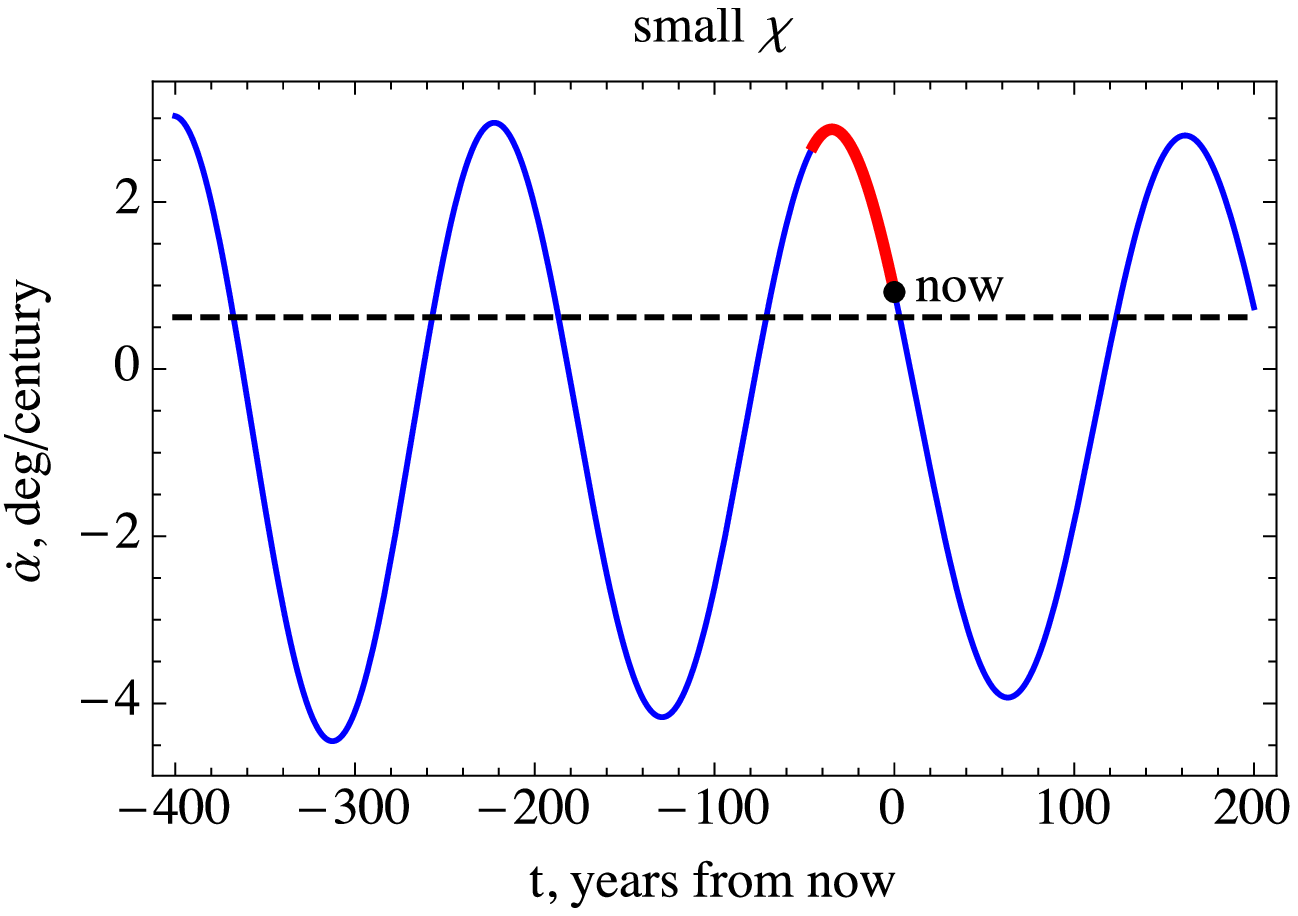}~~\includegraphics[scale=0.45]{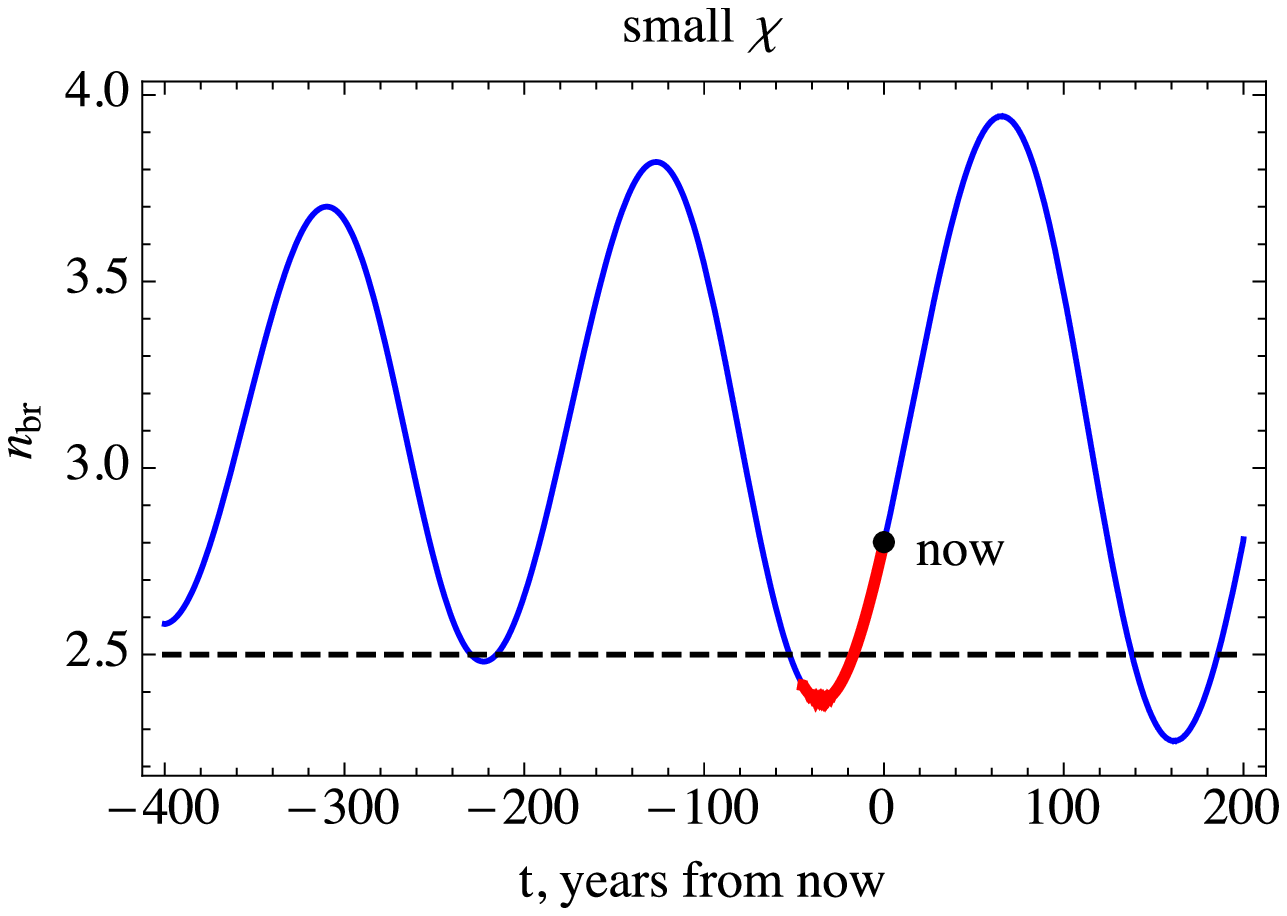}~~\includegraphics[scale=0.45]{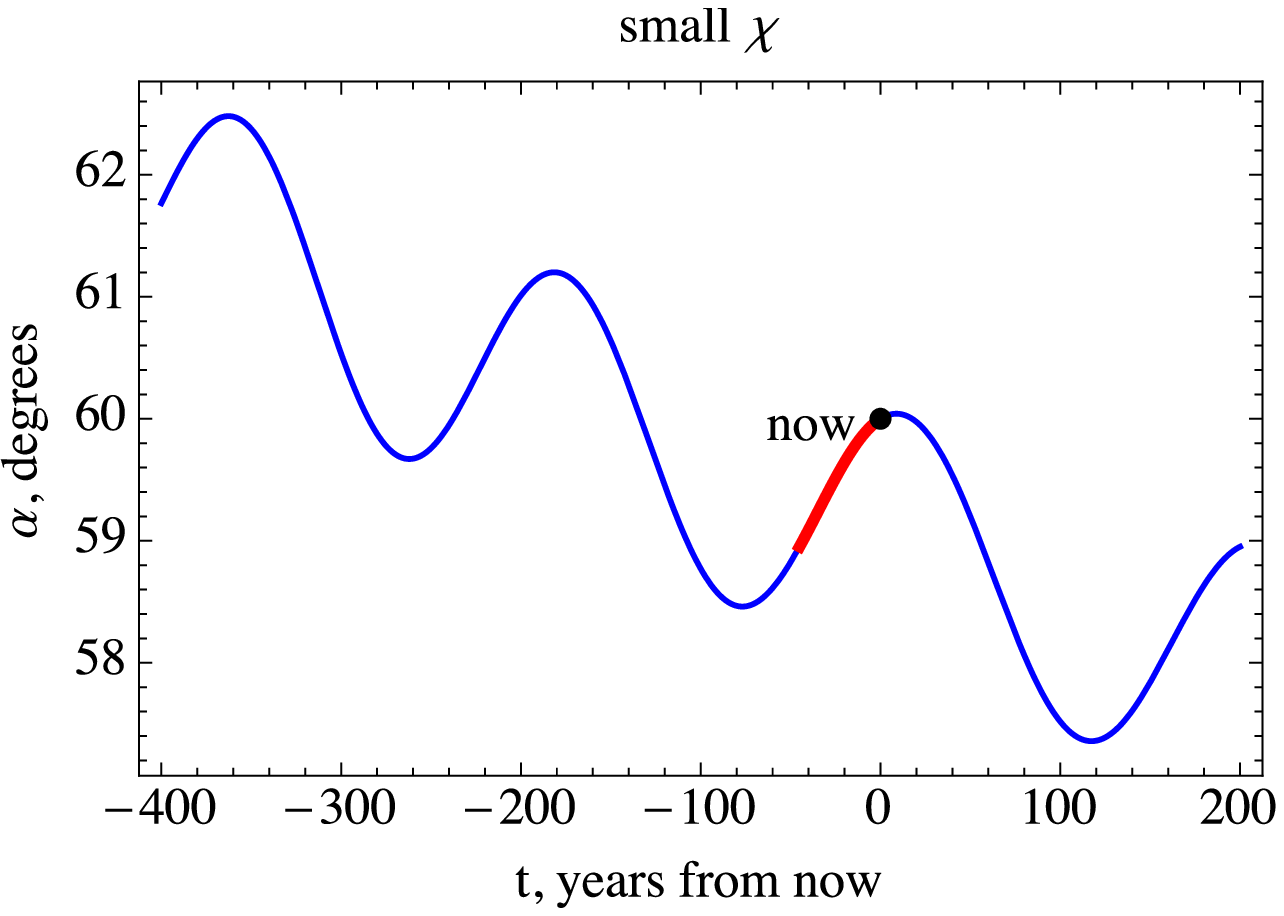}
\caption{Similar to Fig.~\ref{fig:crab1}, but for the \emph{small}
  $\chi$ case. The parameters of this solution at $t=0$ are:
  $\alpha = 60^\circ$, $\theta = 59.1^\circ$, $\chi = 1^\circ$. The
  solution shows order of magnitude agreement with observations. Note
  that for $\chi\ll1$, i.e., when the magnetic and stellar symmetry
  axes are close to each other, $\alpha$ shows a secularly decreasing
  trend, as opposed to the large $\chi$ case, shown in
  Fig.~\ref{fig:crab1}, where $\alpha$ oscillates
  about a constant value.}
\label{fig:crab2}
\end{figure*}

Fig. \ref{fig:crab2} shows the solution for precession model in the case
of small $\chi$. The inferred geometrical parameters are listed in
Table \ref{table:result_crab}. The solution reproduces the observed
data for the past 25 years,  but unlike the case of large $\chi$, the inclination
angle oscillates on top of a global alignment trend. There is the same degeneracy between $\theta$ and $\tau_{\rm prec}$. So, $\sim 100$ years is roughly the smallest period which leads to the good agreement with observations.

\begin{figure*}
\centering
\includegraphics[scale=0.91]{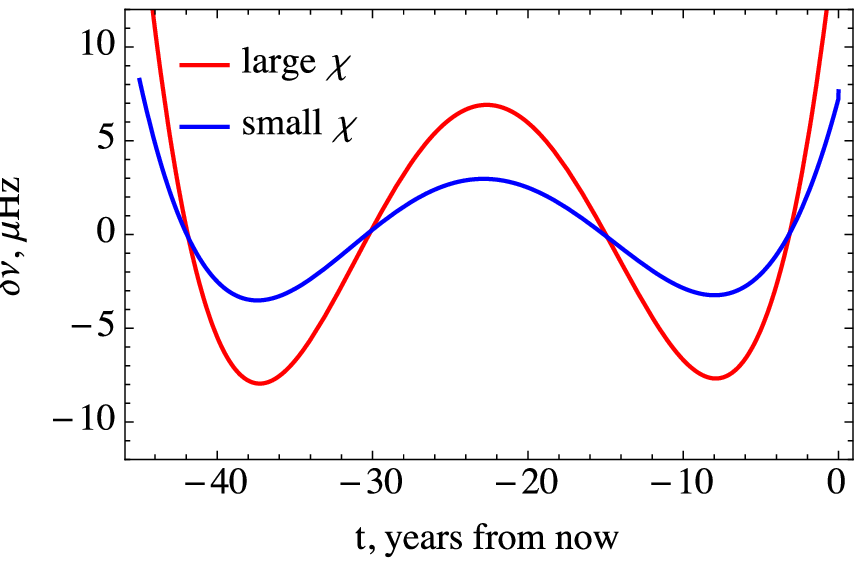}~~~~~\includegraphics[scale=0.6]{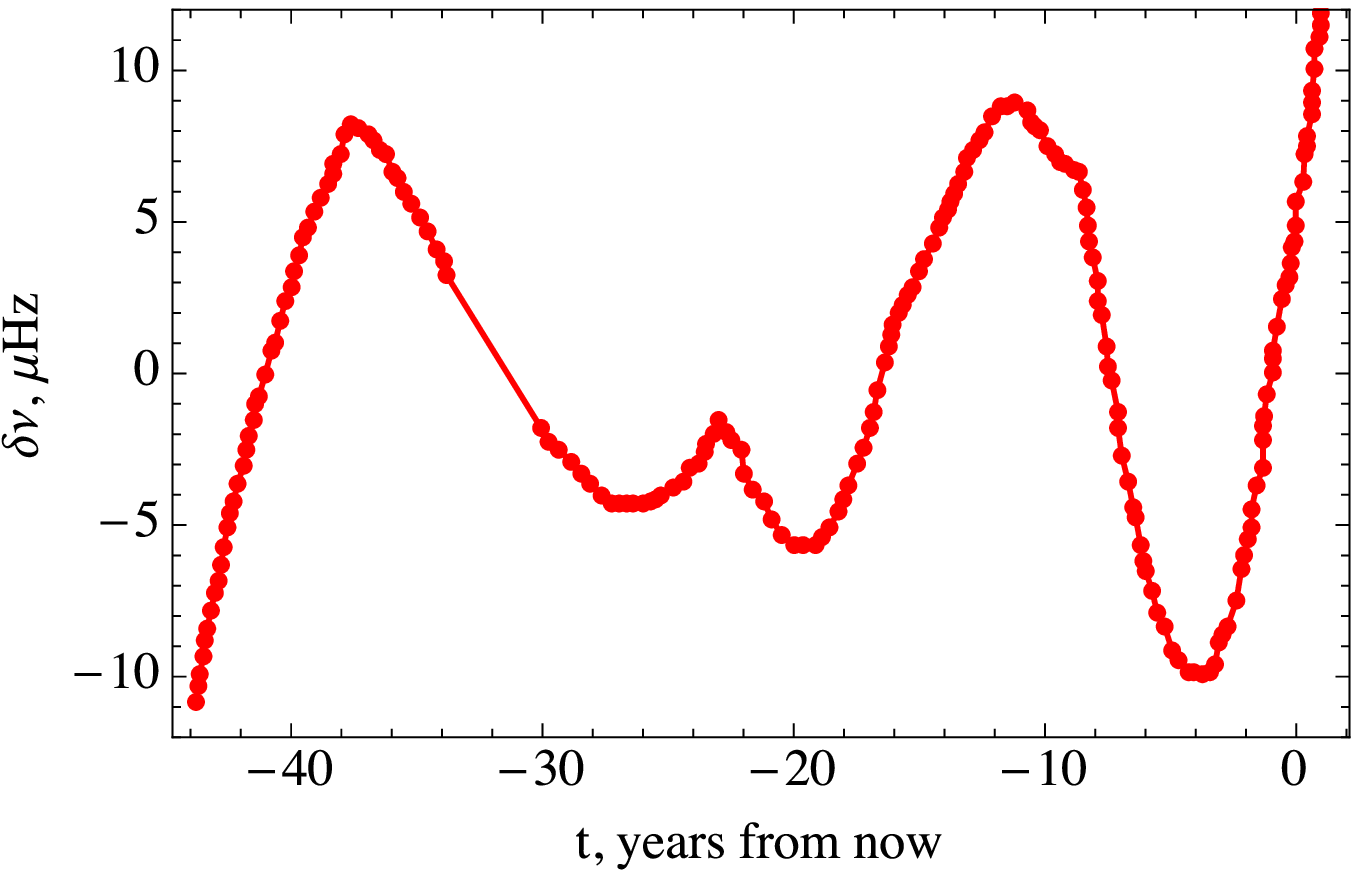}
\caption{Comparison of our model results and the observed frequency
  residual for Crab pulsar. {\bf [Left panel:]} The frequency residuals
  vs time in our numerical simulations are shown with the red line for
  the case of large $\chi$ and the blue line for the case of small
  $\chi$. {\bf [Right panel:]} The observed frequency residual vs
  time. Despite a large number of glitches in the data, our model
  shows an order of magnitude agreement with the observations. Note
  that the variations in the frequency residual occur not on the
  precession period but on a much larger time scale set by the fourth-order
  frequency derivative. In order to observe variations caused by
  precession (as in the case of PSR B1828-11), one needs to observe Crab
  pulsar for at least several precession periods.}
\label{fig:crab1res}
\end{figure*}

\begin{figure*}
\centering
\includegraphics[scale=0.65]{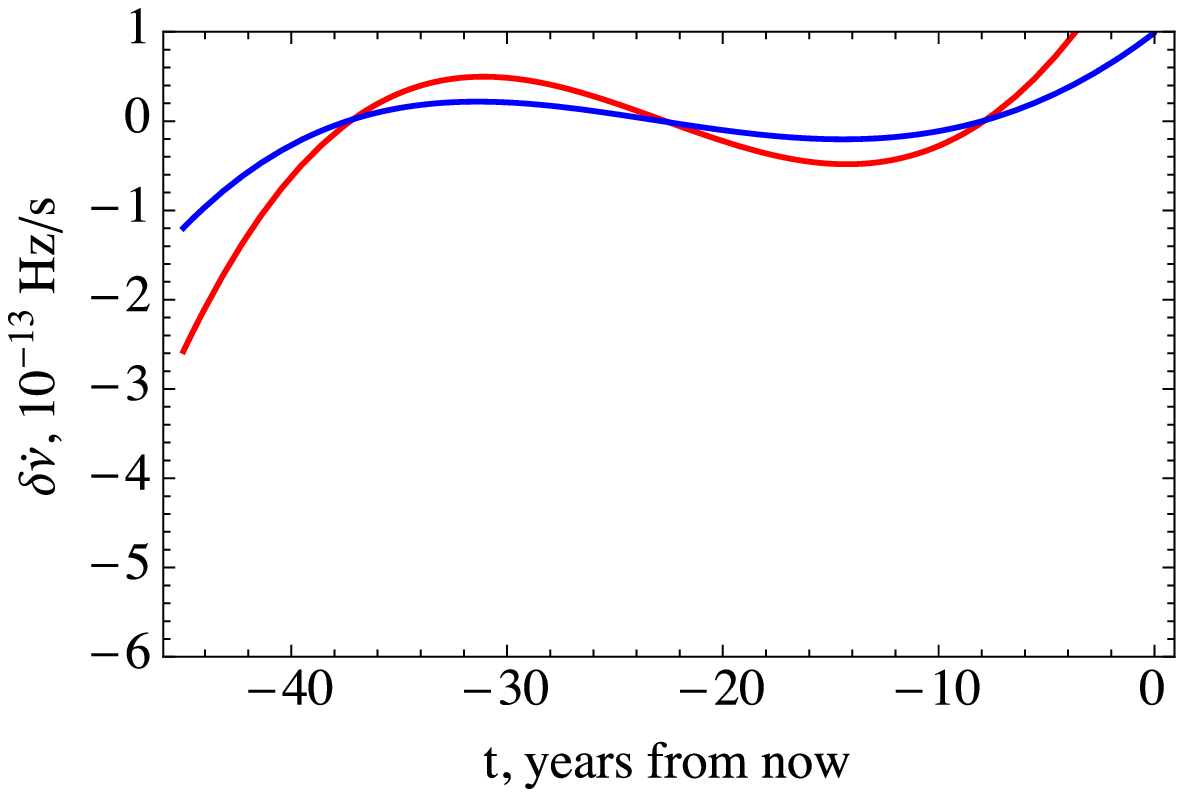}~~~~~\includegraphics[scale=0.62]{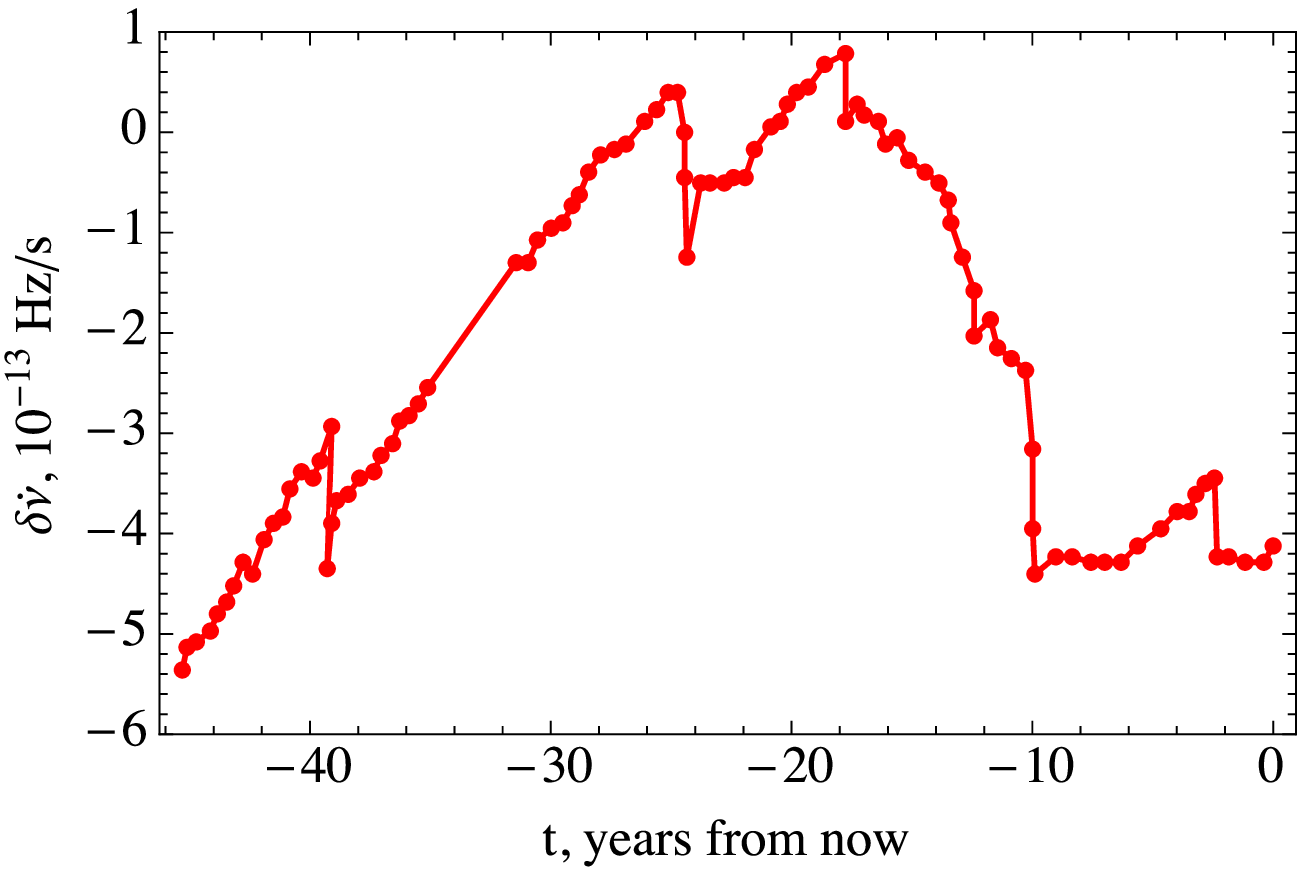}
\caption{The comparison of our Crab modelling results to the observations
  of the frequency derivative residual. {\bf [Left panel]} The red line
  shows the simulated residuals for the
  large $\chi$ case and the blue line for the small $\chi$
  case. {\bf [Right panel]} Observed timing residuals. As in
  Fig.~\ref{fig:crab1res}, our modelling results show an order of
  magnitude agreement with the observed values, as expected
  given the large number of Crab's glitches that are not included into our modelling. }
\label{fig:crab2res}
\end{figure*}

For isolated pulsars it is common to consider a dimensionless
quantity called the \emph{braking index}:
\begin{align}
n_{\rm br} &\equiv \frac{\ddot\Omega \Omega}{\dot\Omega^2}.
\intertext{A spherical pulsar rotating in vacuum has }
n_{\rm br}^{\rm VAC} &= 3 + 2 \tan^{-2} \alpha \ge 3,
\label{eqn:nbr_vac}
\intertext{and a spherical MHD pulsar has}
n_{\rm br}^{\rm MHD} &= 3 + 2 \frac{\sin^2 \alpha\cos^2\alpha}{(1+\sin^2\alpha)^2},~~ 3 \le n_{\rm br}^{\rm MHD}\le 3.25.
\label{eqn:nbr_mhd}
\end{align}
Although $n_{\rm br} \ge 3$ for both models, observations show values
not only less than 3 (e.g., the Crab pulsar has $n_{\rm br} \approx 2.5$)
but even negative values. In fact, braking indices span the range from 
$-10^6$ to $+10^{6}$. If the pulsar experiences precession, it can have an
arbitrary value of braking index both larger and smaller than 3, and
can also be negative. As we show in Figures \ref{fig:crab1} and
\ref{fig:crab2}, the observed braking index of the Crab pulsar can be easily reproduced by our model.

We note that our precessing solutions show that the near-linear
increase of the inclination angle of Crab pulsar will be inevitably
followed by its decrease, and the minimal timescale for this switch is
$\sim 100$ years. It is possible that deviations from the linear
increase could be detectable on much shorter timescales, perhaps as
short as $\sim 20$ years.

For the Crab pulsar, it is quite difficult to fit the timing residuals
data due to the presence of glitches that are not included into our
modelling. Nevertheless, we can compare the characteristic amplitudes
of simulated to observed timing residuals, as shown in
Figs.~\ref{fig:crab1res} and \ref{fig:crab2res}.  In both small and
large $\chi$ cases, our modelling results and observations agree to
within an order of magnitude.

We conclude that precession is a plausible explanation of observed
behaviour of Crab pulsar, quantitatively reproducing the range of observed
frequency derivatives, the braking index,
inclination angle derivative, and the timing residuals.

\section{Discussion}
\label{sect:conclusions}

In this Section, we discuss and compare our results for the 
Crab pulsar and pulsar B1828-11, and discuss astrophysical
implications. 
\subsection{Vacuum vs. Plasma-filled Magnetospheres}
\label{sect:obs_app1}
We start by discussing the differences between the vacuum and MHD
models in terms of their effect on the timing residuals.
As we showed in Sec.~\ref{sect:analytics}, the anomalous torque has no
effect on the timing residuals. This means that the uncertainties in
its estimate do not affect our results. From the point of view of
timing residuals, the difference between vacuum and MHD models is evident in
the denominator of equation (\ref{eq:f}): 
\begin{equation}
\label{eq:f_mhd}
f^{\rm MHD}  =
\frac{\sin2\theta_0\sin2\chi}{4-A(\theta_0,\chi)},
\end{equation}
which for the vacuum case takes the following form:
\begin{equation}
\label{eq:f_vac}
f^{\rm VAC}  =
\frac{\sin2\theta_0\sin2\chi}{2-A(\theta_0,\chi)},
\end{equation}
where $A(\theta,\chi) = 2\cos^2\theta\cos^2\chi+\sin^2\theta\sin^2\chi$.
Since the observed timing residuals, given by
eq.~\eqref{eq:timing_residuals}, are observationally inferred to be
very small, i.e., $\delta \dot P/\dot P \ll 1$, the $f$ factors given
by eqs.~\eqref{eq:f_mhd} and \eqref{eq:f_vac} are also $\ll 1$. To get
a good fit, the values of $\theta$ and $\chi$ must be either near zero $0$ or $90$
degrees. Importantly, the two-peak structure, such as seen in the
residuals of the pulsar B1828-11, is reproduced only for
$g = \tan \chi \tan \theta_0/4 \sim 1$ (see eq.~\ref{eq:g}). So, either $\theta$ is close
to 0 and $\chi$ is close to 90 degrees or vice versa. Thus, the angular
dependence in the denominators of equations (\ref{eq:f_mhd}) and
(\ref{eq:f_vac}) is negligible, i.e., $A\ll 1$, and
\begin{equation}
f^{\rm MHD} = 0.5 f^{\rm VAC}.
\end{equation}

The twice as small residuals implied by plasma-filled models lead to
a larger space of allowed solutions (Sec.~\ref{sect:analytics}) and
allow values of $\chi$ or $\theta$ to be further away from the
extreme values (e.g., $0$ or $90$ degrees) than in the vacuum
models. For instance, previous studies of pulsar B1828-11 in the
context of the vacuum model \citep{2001MNRAS.324..811J, Link} found
$\theta\lesssim1$ degree while our plasma-filled model gives
$\theta = 5$ degrees.

The effect of stellar non-sphericity on the long-term evolution of
pulsar parameters depends on the magnitude of the $\chi$ angle between
the magnetic moment and the principle axis of the star. In the case of large
$\chi$, the angles $\theta$ and $\alpha$ oscillate at the precession
period and their averages show no long-term trends, as seen in
Fig.~\ref{fig:crab1}.

In the case of small $\chi$, the values of $\alpha$ and $\chi$ also
oscillate on the precession timescale but undergo a secular decrease
over longer timescales, as seen in the right panel of
Fig.~\ref{fig:nonspherical}: both vacuum and MHD pulsars evolve toward
alignment.  Thus, if most pulsars are in the small $\chi$ case,
non-sphericity effects will have very few implications for the pulsar
statistics, which favors an alignment trends on the timescale of
$10^6{-}10^7$ years in agreement with spherical pulsar models
(\citealt{tauris98}; \citealt{young};
\citealt{2014MNRAS.443.1891G}). The vacuum pulsars approach the
alignment exponentially fast and the MHD pulsars do so much more
slowly, as a power-law in time. These long-term evolutionary trends
are quantitatively similar to those for spherically-symmetric pulsars
\citep{ptl14}. However, unlike the spherical pulsars, which evolve
toward complete alignment ($\alpha = 0$), non-sphericity causes the
inclination angle to \emph{stabilise} at a finite value,
$\alpha_{\rm stab}>0$.  The precise value of $\alpha_{\rm stab}$ can
differ  between vacuum and MHD models: for instance, for the
parameters chosen in Fig.~\ref{fig:nonspherical}, an MHD pulsar
stabilises at $\alpha^{\rm MHD}_{\rm stab} = 3.46$~deg and a vacuum
pulsar at a much smaller value,
$\alpha^{\rm VAC}_{\rm stab} = 0.35$~deg. Note that since vacuum
pulsars align exponentially fast, they can reach stabilization after
$10^5$ years, which is much shorter than the pulsar lifetime. In
contrast, stabilisation can take $\sim10^8$ years for MHD pulsars, and
it is unclear how many MHD pulsars reach stabilisation in their
lifetime.

\subsection{Crab pulsar vs. B1828-11}
\label{sect:crab_explanations}

We demonstrated that the increase in time of Crab pulsar's inclination
angle could be easily explained if this pulsar experiences precession,
confirming our earlier expectations \citep{ptl14}. A precession
period, $T_{\rm prec} \sim 100$ years, and rotational period,
$P = 0.033$~s, imply a rather small stellar ellipticity
$\varepsilon_{\rm Crab} \lesssim 10^{-11}$ (see
Tab.~\ref{table:result_crab}). Indeed, this value is much smaller than
the minimum allowed value of ellipticity for the pulsar
B1828-11, $\varepsilon_{1828}\sim 10^{-8}$.

When this work was completed, a preprint by \citet{Lai2015} was posted
on the archives. They
considered a precession model for the Crab pulsar in the framework of vacuum
magnetosphere and obtained $\varepsilon = 4 \times 10^{-11}$, slightly
higher than our value. Their geometrical parameters are more extreme,
as one would expect from a vacuum model: $\chi = 0.15$~deg (vs $\ll1$
deg in MHD models) in the
small $\chi$ case and $\theta = 0.1$~deg ($\ll2$ deg) in the large $\chi$ case.

For the Crab pulsar, the MHD spindown formula (\ref{eqn:OmegaDot})
implies a surface magnetic field $B_{12} = 2.33$ for a characteristic
pulsar obliquity angle $\alpha = 60$~deg (the magnetodipole formula
gives a slightly larger value $B_{12} = 3.78$).  The estimate
(\ref{eqn:ell_mag}) for magnetically-induced ellipticity gives
$\varepsilon = 0.6 \times 10^{-11}$, i.e., very close to the inferred
value, so Crab's ellipticity could be accounted for by a
magnetically-induced deformation of its crust.  We note that the we
can only place a rough lower limit on the precession period of the
Crab pulsar, and it could be significantly longer than $100$
years. Long-term observations in the next several decades might help
to break this degeneracy and measure the precession period if it is
not much longer than a century.

The pulsar B1828-11 has approximately the same surface magnetic field,
but its inferred ellipticity is $3$ orders of magnitude larger,
$\epsilon\simeq10^{-8}$.  One way to reconcile this difference from
the Crab pulsar is to
postulate strong ($\sim10^{14}$~G) internal magnetic fields in the
pulsar B1828-11. 
Another possible explanation for the difference in the values of
ellipticities can be found in the framework of rotational ellipticity
model. Equation (\ref{eqn:ell_rot}) gives very large ellipticity
values. However, as \citet{2001MNRAS.324..811J} pointed out, only a
small fraction of this deformation participates in the precession due to
the pinning of magnetic vortexes to the crustal lattice. Crab is
younger, has a lot of glitches, so, the pinning fraction is large and
the effective ellipticity small. Since PSR B1828-11 is older, it is
conceivable that it has a smaller pinning fraction and a larger
effective ellipticity. In this context, it is interesting that all
known pulsars that experience timing noise on the timescale of several
years, have ages larger than that of B1828-11
\citep{2010Sci...329..408L}. Thus, we may expect that the older the
pulsar, the larger the effective ellipticity that participates in
precession.  Dissipative processes in the stellar interior such as
crust-core coupling may also affect the long-term evolution of the neutron
star \citep{crustcore}. 

Pulsar B1828-11 does not show an observed interpulse, and this might seem
at odds with our precessing solution, in which $\alpha$ is very close
to 90 degrees. However, due to the uncertainties in the radio emission
mechanism and the plasma distribution near the polar cap of nearly
orthogonal rotators, it is not clear that an interpulse will
necessarily be visible. It is possible that magnetospheric synchrotron
absorption \citep{Radio} can significantly damp the interpulse
emission. More detailed investigation is
needed to produce reliable models of the observed radio profiles of
this pulsar and independently confirm the geometry inferred from
precession models.

\section{Conclusions}
\label{sec:conclusions}
In this work, we presented the analysis of pulsar evolution that
self-consistently takes into account magnetospheric plasma effects, as
measured in 3D MHD simulations of pulsar magnetospheres. We showed
that the MHD effects expand the allowed parameter space of pulsar
precession solutions and thereby allow for less extreme solutions than
the vacuum models. To facilitate the interpretation of pulsar timing
residual curves, we developed an analytic model that converts the
observed residuals into the pulsar geometrical parameters.

We applied this model to the timing residuals of PSR B1828-11 and obtained
its parameters: e.g., stellar ellipticity and the orientation of the
stellar ellipsoid, magnetic and rotational axes. With the best-fit
parameters, both the period and period derivative timing residual
curves are well-reproduced by our model.  Our best-fitting values for
the pulsar parameters are less extreme than in the vacuum models.

In the context of spherically-symmetric stars, it is expected that the
magnetic and rotational axes of the pulsars evolve toward each other
as the pulsar ages.  Recently a puzzling observation of an opposite
trend was reported for the Crab pulsar \citep{2013Sci...342..598L}. In
this work, we showed that this surprising trend can be explained by
the precession of a neutron star, in agreement with our earlier
expectations (\citealt{ptl14}; see \citealt{Lai2015} for a
consideration of this effect in the vacuum approximation). The data is
consistent with a precession period of the Crab pulsar that is as
short as $\lesssim100$ years. If the true precession period is not
exceedingly larger than that, it could be measured in the next decades,
warranting a continuous monitoring of the Crab pulsar.

{\small
\bibliographystyle{mn2e}
\bibliography{mybib}
}

\appendix

\section{Analytical solution}
\label{sect:appendix}
In this section we describe analytical solution for pulsar
precession. The exact solution could be treated as small perturbations
of free precession motion. We describe free precession in Section
\ref{sect:freePrec}, perturbation caused by anomalous torque are
discussed in Section
\ref{sect:anomPrec}, effects of regular alignment torque are considered in Section
\ref{sect:alignPrec}, two-peak structure in residuals are obtained in Section
\ref{sect:twoPeak} and we finally discuss the effects of the second
ellipticity in Section \ref{sect:secondEll}.

\subsection{Free precession}
\label{sect:freePrec}

For freely precessing body Euler's equations of motion take the
following form:
\begin{equation}
\dot{\boldsymbol L} + {\boldsymbol \Omega} \times \boldsymbol L = 0.
\end{equation}

These equations can be solved analytically (\citet{LLM}) in terms of
Jacobian elliptic functions:
\begin{eqnarray}
L_1/I \Omega_0 &=& \sin\theta_0 \times {\rm cn} (\omega_{\rm p} t, k \tan^2 \theta_0),\\
L_2 /I \Omega_0 &=& \sin\theta_0\sqrt{1+k} \times {\rm sn} (\omega_{\rm p} t, k \tan^2 \theta_0),\\
L_3 /I \Omega_0 &=& \cos\theta_0 \times {\rm dn} (\omega_{\rm p} t, k \tan^2 \theta_0),
\end{eqnarray}
where $k = I_3 (I_2 - I_1)/I_1 (I_3 - I_2) = \varepsilon_{12} (1+
\varepsilon_{13})/(\varepsilon_{13}-\varepsilon_{12}) \approx \varepsilon_{12}/(\varepsilon_{13}-\varepsilon_{12})$ and
$\omega_{\rm p} = \varepsilon_{13} L \cos\theta_0/I_3 \sqrt{1+k}
\approx \varepsilon_{13}\Omega_0 \cos\theta/\sqrt{1+k}$. The precession
period is equal
\begin{equation}
\label{eqA:tauPrec}
\tau_{\rm prec} = \frac{P}{\varepsilon_{13} \cos \theta_0} \frac{2 F(\pi/2,
  k \tan^2\theta_0)}{\pi},
\end{equation}
where $F(\phi, m)$ is the Legendre elliptic integral of the first kind.

In the case of $\varepsilon_{12} = 0$, that is of an axisymmetric star, the
solution takes on a much simpler form:
\begin{eqnarray}
L_1 /I \Omega_0 &=& \sin\theta_0 \times \cos (\omega_{\rm p} t),\\
L_2 /I \Omega_0 &=& \sin\theta_0 \times \sin (\omega_{\rm p} t),\\
L_3 /I \Omega_0 &=& \cos\theta_0, 
\end{eqnarray}
where $\omega_{\rm p} = \varepsilon_{13} \Omega_0 \cos \theta_0$ and $\Omega_0$
is the initial rotational frequency. In this case, the precession
period is simply 
\begin{equation}
\tau_{\rm prec}=P/\varepsilon_{13} \cos \theta_0. \quad {\rm (axisymmetric\ star)}
\label{eqA:tauprecaxisymm}
\end{equation}
\subsection{Effects of the anomalous torque}
\label{sect:anomPrec}

In reality the precession of radio pulsar is not free. In order to
correctly describe it one have to solve the following equation:
\begin{align}
\dot{\boldsymbol L}& + {\boldsymbol \Omega} \times \boldsymbol L =
{\boldsymbol K}=\\
=& K_{\rm aligned} \left(-2 \frac{\boldsymbol
    \Omega}{\Omega}+ \cos \alpha \frac{\boldsymbol
    \mu}{\mu} +\frac{k_3\sin\alpha\cos\alpha}{\Omega R/c} \frac{\boldsymbol
    \Omega\times \boldsymbol \mu}{|\Omega\times\mu|} \right),\nonumber
\end{align}
where we already use MHD torques with $k_0=k_1=k_2=1$.

After we normalize this equation by $I\Omega_0$ and set ${\boldsymbol
  L} = I (\boldsymbol \Omega + \varepsilon_{12}\Omega_2\boldsymbol e_2+ \varepsilon_{12}\Omega_3\boldsymbol e_3)$, we obtain the
following expression:
\begin{align}
\label{eqnA:vecEuler}
\dot{\boldsymbol\omega}& + \frac{1}{\tau_{\rm
  prec}}\left[\frac{\dot\omega_3}{\Omega_0} \boldsymbol e_3+
                         \frac{\varepsilon_{12}}{\varepsilon_{13}}\frac{\dot\omega_2}{\Omega_0}\boldsymbol
                         e_2+\omega_3 \boldsymbol\omega\times\boldsymbol e_3 +
  \frac{\varepsilon_{12}}{\varepsilon_{13}}\omega_3 \boldsymbol\omega\times\boldsymbol e_2\right]
                         =\\
=&
  \frac{\omega^3}{\tau}\left[-2 \frac{\boldsymbol \omega}{\omega} +
  \cos\alpha \frac{\boldsymbol\mu}{\mu} \right]+\frac{\omega^2}{\tau_{\rm
   anom}}\sin \alpha\cos\alpha \frac{\boldsymbol
    \omega\times \boldsymbol \mu}{|\omega\times\mu|},\nonumber
\end{align}
where $\omega_i = \Omega_i/\Omega_0$, $\tau_{\rm prec} =
1/\varepsilon_{13}\Omega_0$, $\tau = I\Omega_0/K_{\rm aligned}^0$ and
$\tau_{\rm anom} =(\Omega_0 R/c) I\Omega_0/k_3K_{\rm aligned}^0$.

In this section we consider $\varepsilon_{12} = 0$ and $\tau_{\rm prec}\ll
\tau_{\rm anom} \ll \tau$. This approximation is
well-satisfied for pulsars and we can treat torque effects as
perturbations of free precession motion.

Using these assumptions and the relation $\cos \alpha = \omega_1
\sin\chi + \omega_3 \cos \chi$ we can write the equations for
perturbations $\delta \omega_i$ in the following way:
\begin{align}
\label{eqnA:sys1}
\delta \dot\omega_1& + \frac{\omega_3 \delta \omega_1 + \omega_1 \delta
  \omega_3}{\tau_{\rm prec}} =\nonumber\\
&= \frac{\omega^2}{\tau_{\rm
  anom}}(\omega_1 \sin\chi + \omega_3 \cos \chi) \omega_2 \cos \chi,\\
\delta \dot\omega_2& - \frac{\omega_3 \delta \omega_2 + \omega_2 \delta
  \omega_3}{\tau_{\rm prec}} =\nonumber\\
\label{eqnA:sys2}
&= \frac{\omega^2}{\tau_{\rm
  anom}}(\omega_1 \sin\chi + \omega_3 \cos \chi) (\omega_3 \sin
  \chi-\omega_1 \cos\chi),\\
\label{eqnA:sys3}
\delta\dot\omega_3& = -\frac{\omega^2}{\tau_{\rm
  anom}}(\omega_1 \sin\chi + \omega_3 \cos \chi) \omega_2 \sin\chi.
\end{align}
Here we neglect the effects of $\tau$ and the first two
terms in brackets in LHS of equation (\ref{eqnA:vecEuler}) as they are
much smaller than $\dot\omega$.

Now we set $\omega_i$ to be equal to their unperturbed values
and solve the system of linear differential equations (\ref{eqnA:sys1})-(\ref{eqnA:sys3}). This is the system of linear differential equations with constant coefficients, so it is straightforward to find a solution:
\begin{align}
\delta \omega_1 =&-\psi_0+\psi_1 \omega_{\rm p} t \sin
                   \omega_{\rm p} t -\psi_2 \cos\omega_{\rm
                   p}t-\nonumber\\
&-\psi_3 \sin^2\omega_{\rm p} t\cos\omega_{\rm p} t, \\
\delta \omega_2 =&-\psi_1 \omega_{\rm p} t \cos
                   \omega_{\rm p} t +\psi_3 \sin\omega_{\rm p} t\cos^2\omega_{\rm p} t \\
\delta \omega_3 =& \psi_0\tan \theta_0\cos\omega_{\rm p}t+\psi_2 \tan\theta_0\cos^2 \omega_{\rm p} t,
\end{align}
where $\omega_{\rm p} = \cos\theta_0/\tau_{\rm prec}$ and
\begin{align}
\psi_0 &=
  \frac{\tau_{\rm prec}}{\tau_{\rm anom}}\sin\chi\cos\chi\cos\theta_0,\\
\psi_1 &=
         \frac{\tau_{\rm prec}}{\tau_{\rm anom}}\left(\cos^2\chi\sin\theta_0-\sin^2\chi\frac{2\sin\theta_0-\sin^3\theta_0}{4\cos^2\theta_0}\right),\\
\psi_2 &= \frac{\tau_{\rm prec}}{\tau_{\rm anom}}\sin^2\chi\sin\theta_0/2,\\
\psi_3
       &=\frac{\tau_{\rm prec}}{\tau_{\rm anom}}\frac{\sin^2\chi}{4}\frac{\sin^3\theta_0}{\cos^2 \theta_0}.
\end{align}

It is important to notice that the amplitude of $\omega$ remains
unperturbed: $\omega_1 \delta\omega_1 + \omega_2\delta\omega_2 +
\omega_3 \delta\omega_3 \equiv 0$. This is because the anomalous
torque is perpendicular to $\boldsymbol \Omega$ and thus cannot change
its amplitude. 

On the other hand, anomalous torque causes the changes in $\theta$
since $\delta \omega_3 = - \sin\theta_0 \delta \theta$.
\begin{equation}
\label{eqnA:theta_var}
\delta\theta = - \frac{\cos \omega_{\rm p}t}{\cos\theta_0} (\psi_0 +
\psi_2 \cos \omega_{\rm p}t).
\end{equation}
The maximum deviation is given by
\begin{equation}
\max|\theta - \theta_0|\approx  \frac{\tau_{\rm prec}}{\tau_{\rm anom}}\frac{\sin2\chi+\sin^2\chi\tan\theta_0}{2}.
\end{equation}

\begin{figure*}
\centering
\includegraphics[scale=0.45]{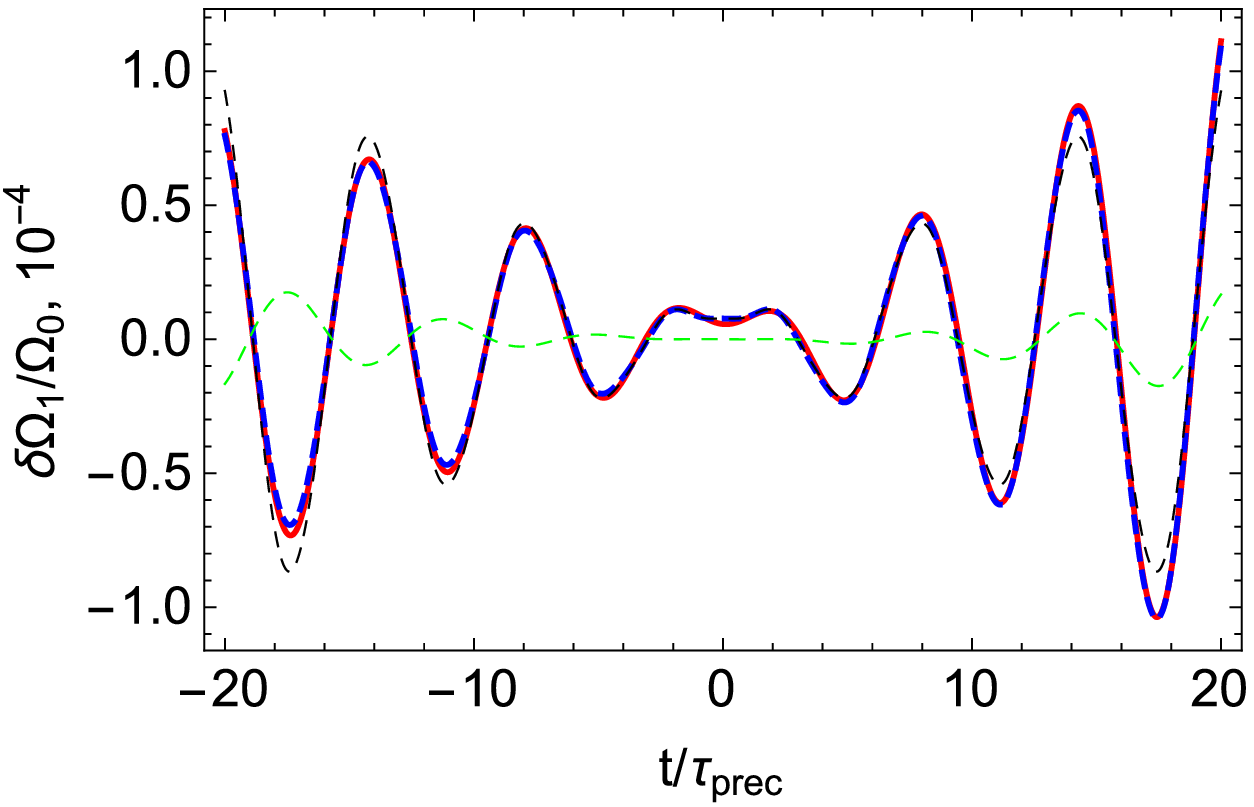}~~\includegraphics[scale=0.45]{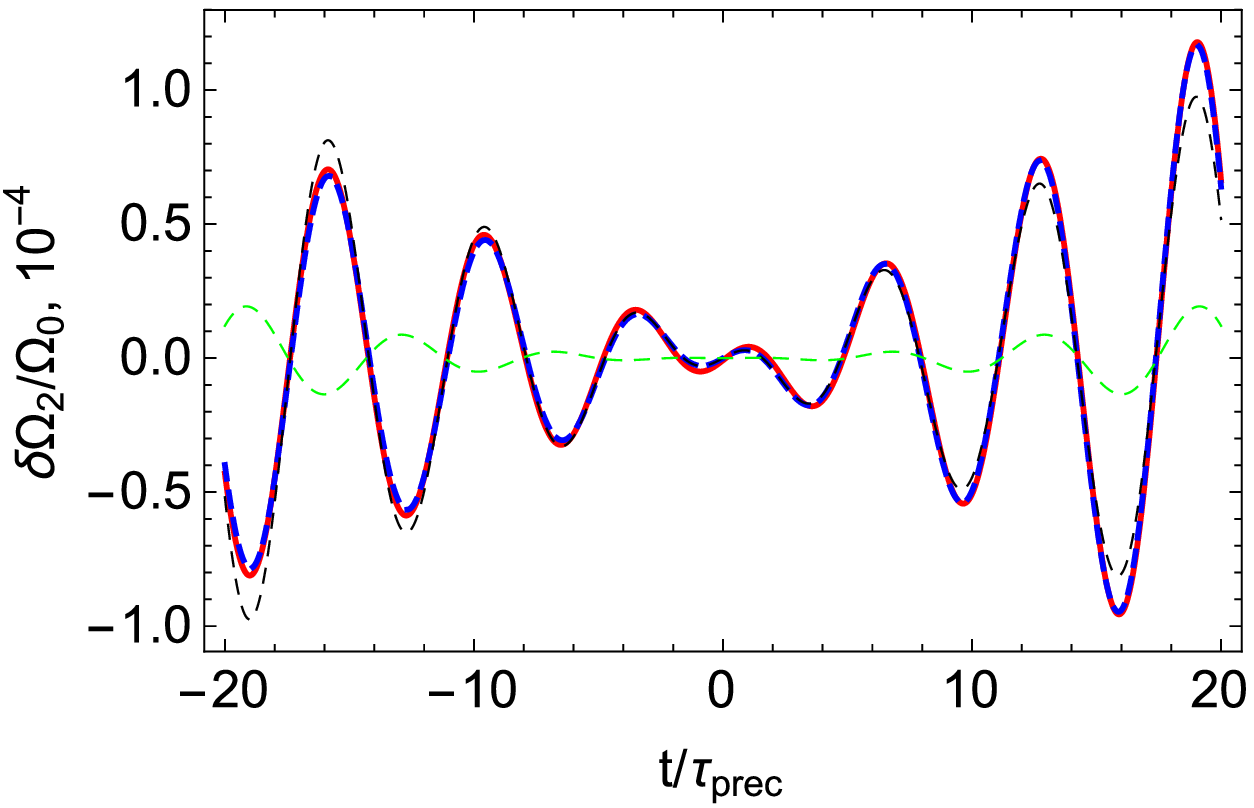}~~\includegraphics[scale=0.61]{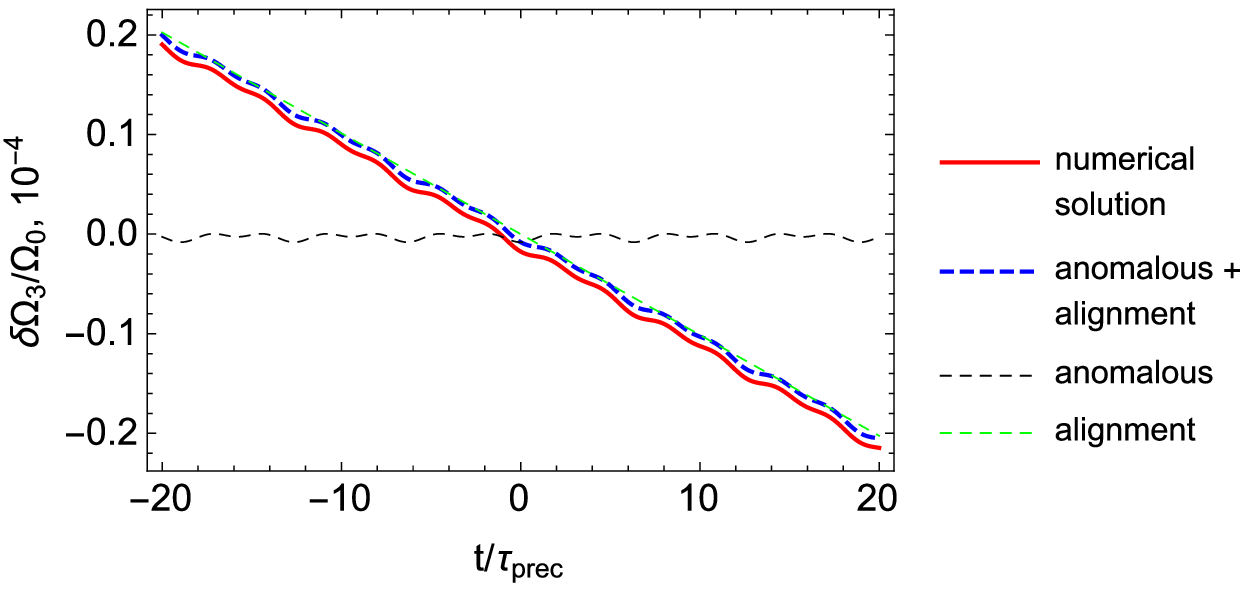}
\caption{The comparison of exact numerical solution with analytical
  asymptotic for $\chi = 88.5$ deg, $\theta_0 =
  6$ deg. The figure shows an excellent agreement of numerical solution and observations. This allows us to use simple analytical expressions (\ref{eqn:result})--(\ref{eqn:resultT}) instead of numerical procedures.}
\label{fig:om_var}
\end{figure*}

\subsection{Effects of the alignment torque}
\label{sect:alignPrec}
If one considers in equation (\ref{eqnA:vecEuler}) the contribution
of alignment torque, it is possible to complete exactly the same
procedure as in Section \ref{sect:anomPrec}. The solution in this case is quite different:
\begin{align}
\delta\omega_1 &= \phi_2 \sin 2\omega_{\rm p}t - \phi_3 \omega_{\rm p}t
                 \cos \omega_{\rm p}t +\phi_4 \omega_{\rm p}^2t^2 \sin\omega_{\rm p} t,\\
\delta\omega_2 &= \phi_0+\phi_1\cos \omega_{\rm p}t-\phi_2 \cos 2
                 \omega_{\rm p}t - \phi_3 \omega_{\rm p}t \sin
                 \omega_{\rm p}t- \nonumber\\
 &- \phi_4 \omega_{\rm p}^2t^2\cos \omega_{\rm p}t,\\
\delta\omega_3 &= (\phi_2 \sin \omega_{\rm p} t -\phi_4 \omega_{\rm p}
                 t)\cot \theta_0,
\end{align}
where
\begin{align}
\phi_0 &=\frac{\tau_{\rm prec}}{\tau}
         \frac{2-\tan^2\theta_0}{2}\sin\chi \cos\chi,\\
\phi_1 &=\frac{\tau_{\rm prec}}{\tau} \frac{\sin^2\chi\tan\theta_0}{2},\\
\phi_2 &=\frac{\tau_{\rm prec}}{\tau} \sin\chi\cos\chi \tan^2\theta_0,\\
\phi_3 &=\frac{\tau_{\rm prec}}{\tau}\frac{7+\cos2\chi}{4}\tan\theta_0,\\
\phi_4 &=\frac{\tau_{\rm prec}}{\tau} \frac{3-\cos
         2\chi}{4} \tan \theta_0.
\end{align}

If we combine the effects of both alignment and anomalous torque we
obtain the total perturbation. Fig. \ref{fig:om_var} compares exact
numerical solution with total perturbation as well as the
perturbations caused by alignment and anomalous torques separately.

The alignment torque causes $\omega$ to change. The perturbation to
the frequency takes the following form:
\begin{align}
\delta \omega& = \omega_1 \delta \omega_1 + \omega_2 \delta \omega_2 +
\delta \omega_3 \omega_3 = \\
&=-\frac{\tau_{\rm prec}}{\tau} \omega_{\rm p}t (3-\cos
  2\chi+7\tan^2\theta_0-\cos2\chi\tan^2\theta_0)\cos\theta_0/2+\nonumber\\
&+\frac{\tau_{\rm prec}}{\tau}\sin2\chi\sin\theta_0 (\sin
  \omega_{\rm p}t + 1/8 \tan\theta_0\tan\chi\sin 2 \omega_{\rm p}t).\nonumber
\end{align}
Here we can see spin down as well as the two harmonics that are
actually presented in observations.

Summing up, the observed residual curves are the first-order perturbation over free precession motion caused by the alignment torque. The first-order residual has up to two harmonics. 

\subsection{Two-peak structure}
\label{sect:twoPeak}

The frequency residual $\delta \omega$ could be found in much more
simple way. In order to find it we multiply equation
(\ref{eqnA:vecEuler}) by $\boldsymbol \omega$:
\begin{equation}
\label{eqnA:resOmega}
\dot\omega =- \frac{\omega_3 \dot \omega_3+\varepsilon_{12}/\varepsilon_{13}\omega_2
\dot \omega_2}{\omega \tau_{\rm prec} \Omega_0} -\frac{\omega^3 (1+\sin^2\alpha)}{\tau} 
\end{equation}

Puting zero-order solution for $\varepsilon_{12}=0$ in RHS of this equation
gives the following equation:
\begin{align}
\dot\omega &= -\frac{1}{\tau}
(2-\cos^2\theta_0\cos^2\chi-1/2\sin^2\theta_0\sin^2\chi)+\\
&+\frac{1}{\tau}\frac{\sin 2\theta_0\sin 2\chi}{2} (\cos \omega_{\rm p} t +\cos 2\omega_{\rm p} t
\tan\theta_0\tan\chi/4).
\end{align}

The first term of this expression is the observed spindown. Having
that we can rewrite this equation in the following way:
\begin{align}
\dot\omega=-\frac{\dot P^{\rm obs}}{ P^{\rm obs}}\left(1- \frac{\sin 2\theta_0\sin 2 \chi  (\cos\omega_{\rm
  p}t + \cos 2\omega_{\rm p} t\tan\theta_0\tan\chi/4)}{4 - 2
  \cos^2\theta_0 \cos^2 \chi - \sin^2\theta_0\sin^2\chi}\right).
\end{align}

Finally, the residual in $\dot P$ now takes a form
\begin{equation}
\delta \dot P = - \dot P^{\rm obs} f(\theta_0,\chi) (\cos\omega_{\rm
  p}t+g(\theta_0,\chi) \cos 2\omega_{\rm p}t)
\end{equation} 
with 
\begin{equation}
\label{eqA:f}
f(\theta_0,\chi) =
\frac{\sin2\theta_0\sin2\chi}{4-2\cos^2\theta_0\cos^2\chi-\sin^2\theta_0\sin^2\chi},
\end{equation}
\begin{equation}
\label{eqA:g}
g(\theta,\chi)=\tan\theta_0\tan\chi/4.
\end{equation}

The amplitude of residual is proportional to $\sin 2\theta_0 \sin 2
\chi$. If we have residuals which are much smaller than $\dot P^{\rm
  obs}$ (which is usually the case), $\theta_0$ and $\chi$ should be
close either to 0 or to 90 deg. As observed residuals are much smaller
than $\dot P^{\rm obs}$ and we observe two-bump structure (so,
$g(\theta_0,\chi)\approx 1$), one of the angles should be very close to
0, and another one -- to 90 deg.

\subsection{Effects of the second ellipticity}
\label{sect:secondEll}

For $\varepsilon_{12},\varepsilon_{23}\ll1$, one can show that
\begin{equation}
\frac{\varepsilon_{12}}{\varepsilon _{13}} \approx \frac{k}{k+1}.
\end{equation}
In this case, zero-order solution implies
\begin{equation}
\omega_3 \dot \omega_3 + \varepsilon_{12}/\varepsilon_{23} \omega_2 \dot \omega_2 \equiv 0.
\end{equation}
The solution for perturbations takes now very similar form:
\begin{equation}
\label{eqnA:result}
\delta \dot P = - \dot P^{\rm obs} f(\theta_0,\chi) \mathcal{P}(\theta_0, \chi,t)
\end{equation}
with
\begin{align}
\mathcal{P} &= {\rm cn}(\omega_p
  t)  {\rm dn}(\omega_p
  t) +  {\rm cn}(2\omega_p
  t) \frac{2 g(\theta_0,\chi)}{(1+ {\rm dn}(2\omega_p
  t))} +\\
&+( {\rm dn}(2\omega_p
  t)-1) \left(\frac{g(\theta_0,\chi)}{(1+ {\rm dn}(2\omega_p
  t))} + \frac{ g(\theta_0,\chi)^{-1}}{8(1+ {\rm cn}(2\omega_p
  t))}\right).\nonumber
\end{align}

It shows that the second ellipticity does not qualitatively change the behaviour of solution and only influences the precession period (see eq. (\ref{eqA:tauPrec})).

\bsp

\label{lastpage}

\end{document}